\documentclass[
reprint,
amsmath,
amssymb,
aps,
floatfix,
]{revtex4-2}

\usepackage{graphicx}
\usepackage{cprotect}
\usepackage[export]{adjustbox}

\makeatletter
\@ifclassloaded{revtex4-2}{%
  \let\old@makecaption=\@makecaption%
  \usepackage{subcaption}%
  \let\@makecaption=\old@makecaption%
}{%
  \usepackage{subcaption}
}
\makeatother

\usepackage{dcolumn}
\usepackage{booktabs}
\usepackage{array}
\usepackage{multirow}
\usepackage{tabularx}
\usepackage{enumitem}
\usepackage{paralist}

\usepackage{bm}
\usepackage{siunitx}
\DeclareSIUnit\angstrom{\text{Å}}
\usepackage{braket}
\usepackage{mathtools}

\usepackage{amsfonts}
\usepackage{amsthm}

\usepackage{hyperref}
\usepackage{url}
\usepackage{spverbatim}

\usepackage{printlen}
\usepackage{xspace}

\usepackage[capitalize]{cleveref}
\usepackage{xr}

\usepackage{relsize}
\newcommand{\smallunderscore}{\textscale{.5}{\textunderscore}}

\usepackage{derivative}

\usepackage{mhchem}

\usepackage{ifthen}
\newcommand{\gp}[2][]{%
\ifthenelse{\equal{#1}{}}%
{\color{ForestGreen}#2}%
{\color{ForestGreen}[GP: #1] #2}%
}
\newcommand{\jq}[2][]{%
\ifthenelse{\equal{#1}{}}%
{\color{orange}#2}%
{\color{orange}[JQ: #1] #2}%
}

\makeatletter
\DeclareRobustCommand\onedot{\futurelet\@let@token\@onedot}
\def\@onedot{\ifx\@let@token.\else.\null\fi\xspace}

\def\eg{e.g\onedot}

\def\ie{i.e\onedot}

\def\wrt{w.r.t\onedot}

\makeatother

\newcommand{\QE}{\texttt{Quantum ESPRESSO}\xspace}
\newcommand{\PtoW}{\texttt{pw2wannier90.x}\xspace}
\newcommand{\WAN}{\texttt{Wannier90}\xspace}
\newcommand{\OMX}{\texttt{OpenMX}\xspace}
\newcommand{\AiiDA}{\texttt{AiiDA}\xspace}

\newcommand{\schrodinger}{Schr\"{o}dinger\xspace}
\newcommand{\lowdin}{L\"{o}wdin\xspace}

\DeclareMathOperator\erfc{erfc}

\newcommand{\densitymatrix}{$\braket{\mathbf{r} | P_{\mathbf{k}} | \mathbf{r}^{\prime}}$\xspace}

\newcommand{\pmin}{$p_{\min}$\xspace}
\newcommand{\pmax}{$p_{\max}$\xspace}

\newcommand{\ef}{$E_F$\xspace}
\newcommand{\dwfc}{$d_{\text{WFC}}$\xspace}
\newcommand{\omegawf}{$\Omega_{\text{WF}}$\xspace}

\newcommand{\kpt}{$k$-point\xspace}
\newcommand{\kpts}{\kpt{}s\xspace}

\newcommand{\SrVOthree}{\ce{SrVO3}\xspace}

\newcommand{\AutwoTi}{\ce{Au2Ti}\xspace}
\newcommand{\AlthreeV}{\ce{Al3V}\xspace}
\newcommand{\BasixGeten}{\ce{Ba6Ge10}\xspace}
\newcommand{\BrtwoTi}{\ce{Br2Ti}\xspace}

\newcommand{\epfl}{Theory and Simulations of Materials (THEOS), and National Centre
for Computational Design and Discovery of Novel Materials (MARVEL), \'Ecole
Polytechnique F\'ed\'erale de Lausanne, 1015 Lausanne, Switzerland}
\newcommand{\psich}{Laboratory for Materials Simulations (LMS),
Paul Scherrer Institut (PSI), CH-5232 Villigen PSI, Switzerland}

\begin{document}

\title{Projectability
  disentanglement for accurate and automated electronic-structure Hamiltonians}

\author{Junfeng Qiao}
\email{junfeng.qiao@epfl.ch}
\affiliation{\epfl}
\author{Giovanni Pizzi}
\affiliation{\epfl}
\affiliation{\psich}
\author{Nicola Marzari}
\affiliation{\epfl}
\affiliation{\psich}

\date{\today}

\begin{abstract}
  Maximally-localized Wannier functions (MLWFs) are a powerful and broadly used
    tool to characterize the electronic structure of materials, from chemical
    bonding to dielectric response to topological properties.
  Most generally, one can construct MLWFs that describe isolated band manifolds,
    \eg for the valence bands of insulators, or entangled band manifolds, \eg in
    metals or describing both the valence and the conduction manifolds in
    insulators.
  Obtaining MLWFs that describe a target manifold accurately and with the most
    compact representation often requires chemical intuition and trial and error, a
    challenging step even for experienced researchers and a roadblock for automated
    high-throughput calculations.
  Here, we present a powerful approach that automatically provides MLWFs spanning
    the occupied bands and their natural complement for the empty states, resulting
    in Wannier Hamiltonian models that provide a tight-binding picture of optimized
    atomic orbitals in crystals.
  Key to the success of the algorithm is the introduction of a projectability
    measure for each Bloch state onto atomic orbitals (here, chosen from the
    pseudopotential projectors) that determines if that state should be kept
    identically, discarded, or mixed into a disentangling algorithm.
  We showcase the accuracy of our method by comparing a reference test set of 200
    materials against the selected-columns-of-the-density-matrix algorithm, and its
    reliability by constructing Wannier Hamiltonians for \num{21737} materials from
    the Materials Cloud.
\end{abstract}

\maketitle

\section{Introduction}
In periodic crystals, the electronic structure is usually described using
  one-particle Bloch wavefunctions.
While choosing a basis set that is also periodic to describe these
  wavefunctions can often be beneficial, an alternative approach is to adopt
  localized orbitals in real space.
One such choice of orbitals are Wannier functions (WFs), that can be obtained
  by Fourier transforming the periodic wavefunctions from reciprocal to real
  space.
WFs are not unique, as they depend on the choice of the gauge (\ie, the choice
  of the phases of the wavefunctions) at each point in the Brillouin zone (BZ).
Maximally-localized Wannier functions (MLWFs) \cite{Marzari1997,Souza2001,
    Marzari2012,Pizzi2020} are obtained by a gauge choice that is optimized to
  provide the most localized set of WFs, \ie, those that minimize the sum of
  their quadratic spread in real space \cite{Marzari1997}.
Having a very localized representation of the electronic structure not only
  provides an insightful analysis of chemical bonding in solids, but also brings
  a formal connection between the MLWF centers and the modern theory of electric
  polarization \cite{Resta2007}.
Moreover, the real-space locality of MLWF allows for accurate and fast
  interpolation of physical operators \cite{Lee2005}, enabling calculations of
  material properties that require dense samplings of the BZ, such as Fermi
  surface, orbital magnetization \cite{Lopez2012}, anomalous Hall conductivity
  \cite{Wang2006,Yates2007}, and spin Hall conductivity \cite{Qiao2018}, to name
  a few.
Practically, one obtains MLWFs starting from a set of Bloch wavefunctions,
  calculated \eg, from density-functional theory (DFT).
Often, these Bloch states are projected onto some localized orbitals (usually
  chosen by the user) to generate initial guesses for MLWFs.
In an insulator, by minimizing the spread functional \cite{Marzari1997} which
  measures localization, one can obtain a set of MLWFs, \ie, ``Wannierize'' a
  material.
The Wannierization contains an additional disentanglement step \cite{Souza2001}
  if the target Bloch states are not isolated from other band manifolds.
For such entangled bands---metals or the conduction bands of insulators---one
  needs to first identify the relevant Bloch states that will be used to
  construct MLWFs, and then mix or ``disentangle'' these from all the Bloch
  states \cite{Souza2001}.
Practically, the choices for the initial projections and states to be
  disentangled substantially influence the shape and the quality of the final
  MLWFs.

In recent years, a lot of effort has been devoted to obtaining high-quality
  MLWFs and automate the Wannierization procedure.
Focus of the research can be categorized into the following classes:
  \begin{inparaenum}[(a)] \item Novel minimization algorithms, such as: the
  symmetry-adapted WF method that adds constraints to impose the symmetries of
  the resulting WFs \cite{Sakuma2013}; the simultaneous diagonalization algorithm
  that directly minimizes the spread functional for an isolated (or
  ``$\Gamma$-only'') system \cite{Gygi2003}; the partly-occupied WF method, where
  the total spread is directly minimized in one step
  \cite{Thygesen2005,Thygesen2005a}, rather than performing a two-step
  minimization for its gauge-invariant and gauge-dependent parts as in the
  standard procedure \cite{Souza2001}; or the variational formulation, that
  combines single-step optimization with manifold optimization to make the
  minimization algorithm more robust \cite{Damle2019}; \item new forms for the
  spread functional, such as the selectively localized WFs (SLWFs) for which only
  a subset of WFs of interest are localized and a penalty term is added to
  constrain the position of the WF centers \cite{Wang2014}, or the
  spread-balanced WF method, that adds a penalty term to distribute the spread as
  uniformly as possible among all WFs \cite{Fontana2021}; \item targeting a
  subset of orbitals, \eg SLWF for a subset of MLWFs \cite{Wang2014} or the
  optimized projection functions method where starting projections for the
  Wannierization are generated from a larger group of initial ones
  \cite{Mustafa2015}; \item matrix manifold algorithms instead of projection
  methods to construct a smooth gauge in a non-iterative way
  \cite{Cances2017,Gontier2019}; \item basis-vector decomposition of the density
  matrix, \eg the selected columns of the density matrix (SCDM) algorithm
  \cite{Damle2015,Damle2018}, that starts from the density matrix of the system
  and uses QR decomposition with column pivoting (QRCP) to automatically generate
  an optimal set of basis vectors from the columns of the density matrix.
\end{inparaenum}

At the same time, high-throughput (HT) calculations have become increasingly
  popular for materials discovery and design.
Calculations and results managed by workflow engines are collected into
  databases of original calculations, such as the Materials Project
  \cite{Jain2013}, AFLOW \cite{Curtarolo2012}, OQMD \cite{Saal2013}, CMR
  \cite{CMR}, and the Materials Cloud \cite{Talirz2020}, or aggregated, as in
  NOMAD \cite{Draxl2018}.
Thanks to recent research advances on Wannierization algorithms, it starts now
  to be possible to run HT Wannierizations for many materials and generate
  tight-binding (TB) models that reliably describe their physics.
So far, several attempts have been made in this direction.
\citet{Gresch2018} gathered \si{195} Wannier TB Hamiltonians and applied
  post-processing symmetrization to study strained III-V semiconductor materials.
\citet{Vitale2020} implemented the SCDM algorithm and designed a protocol to
  determine automatically the remaining free parameters of the algorithm; this
  protocol, implemented into automated workflows, was verified to work well for
  band interpolations on a set of 200 structures (metals, or valence and
  conduction bands of insulators) and 81 insulators (valence bands only).
\citet{Garrity2021} accumulated a Wannier TB Hamiltonian database of \si{1771}
  materials using the standard hydrogenic orbital projections.
However, there are still several challenges for an accurate and automated HT
  Wannierization, some of which might be more relevant depending on the research
  goal and the specific property to compute: MLWFs should be able to faithfully
  represent the original band structure, often (\eg, for transport properties) at
  least for those bands close to the Fermi energy; MLWFs should resemble the
  physically intuitive atomic orbitals for solids that would enter into Bloch
  sums; the algorithm should be fully and reliably automated and the
  implementation should be efficient for HT calculations.

To overcome the challenges mentioned above, in this paper we present a new
  methodology for automated Wannierization.
First, we choose physically-inspired orbitals as initial projectors for MLWFs,
  that is, the pseudo-atomic orbitals (PAOs) from pseudopotentials
  \cite{Agapito2016}.
Then, for each state $\ket{n \mathbf{k}}$ ($n$ is the band index, $\mathbf{k}$
  is the Bloch quasi-momentum) we decide if it should be dropped, kept
  identically, or thrown into the disentanglement algorithm depending on the
  value of its projectability onto the chosen set of PAOs, replacing the standard
  disentanglement and frozen manifolds based only on energy windows.
This approach naturally and powerfully targets the TB picture of atomic
  orbitals in crystals, as it will also become apparent from our results.
Moreover, we fully automate this approach and implement it in the form of
  open-source \AiiDA \cite{Pizzi2016,Huber2020,Uhrin2021} workflows.
To assess its effectiveness and precision, we compare the quality of the band
  interpolation and the locality of the Wannier Hamiltonians generated with the
  present approach, which we name as projectability-disentangled Wannier
  functions (PDWFs), with the results from the SCDM algorithm \cite{Vitale2020}.
Statistics from 200 materials demonstrate that PDWFs are more localized and
  more atomic-like, and the band interpolation is accurate at the meV scale.
Furthermore, to demonstrate the reliability and automation of our method and
  workflows, we carry out a large-scale high-throughput Wannierization of
  \num{21737} materials from the Materials Cloud\cite{Talirz2020,MC3D}.

To set the context for the following paragraphs, here we briefly summarize the
  notations for WFs; a detailed description can be found in
  Refs.~\cite{Marzari1997,Souza2001,Marzari2012}.
WFs $\ket{ w_{n \mathbf{R}} }$ are unitary transformations of Bloch
  wavefunctions $\ket{\psi_{m\mathbf{k}}}$, given by
  \begin{equation}
    \ket{w_{n
          \mathbf{R}}} = \frac{V}{(2 \pi)^{3}} \int_{\mathrm{BZ}} \mathrm{d} \mathbf{k}
    \mathrm{e}^{-\mathrm{i} \mathbf{k} \cdot \mathbf{R}} \sum_{m=1}^{J
    \operatorname{or} J_{\mathbf{k}}} \ket{\psi_{m \mathbf{k}} } U_{m n
        \mathbf{k}},
  \end{equation}
  where $\mathbf{k}$ and $\mathbf{R}$ are the Bloch
  quasi-momentum in the BZ and a real-space lattice vector, respectively; $m$ is
  the band index, and $n$ is the Wannier-function index (running from 1 to the
  number of WFs $J$).
For an isolated group of bands, $J$ is equal to the number of bands, and the
  $U_{m n \mathbf{k}}$ are unitary matrices; for entangled bands, the number of
  bands considered at each $k-$point is $J_{\mathbf{k}} \ge J$, and the $U_{m n
        \mathbf{k}}$ are semi-unitary rectangular matrices.
MLWFs are the minimizers of the quadratic spread functional \cite{Marzari1997}
  \begin{equation} \label{eq:mv_spreads} \Omega=\sum_{n=1}^{J}\left[\braket{ w_{n
        \mathbf{0}}|\mathbf{r}^2| w_{n \mathbf{0}}}-\left|\braket{ w_{n
        \mathbf{0}}|\mathbf{r}| w_{n \mathbf{0}}}\right|^{2}\right].
\end{equation}
Since \cref{eq:mv_spreads} is a minimization problem with multiple local
  minima, initial guesses for $U_{m n \mathbf{k}}$ substantially influence the
  optimization path and the final minimum obtained.
In order to target the most localized and chemically appealing solution,
  \citet{Marzari1997} used hydrogenic wavefunctions $\ket{g_n}$ (\ie, analytic
  solutions of the isolated hydrogenic \schrodinger equation) to provide a set of
  sensible initial guesses $\ket{\phi_{n \mathbf{k}}}$, after projection on the
  space defined by the relevant Bloch states: \begin{equation}
  \label{eq:wan_init_proj} \ket{ \phi_{n \mathbf{k}} } = \sum_{m=1}^{J
  \operatorname{or} J_{\mathbf{k}}} \ket{\psi_{m \mathbf{k}}} \braket{\psi_{m
      \mathbf{k}} | g_{n} }.
\end{equation}
The projection matrices $A_{mn\mathbf{k}} = \braket{\psi_{m \mathbf{k}} |
      g_{n}}$, after \lowdin orthonormalization \cite{Loewdin1950}, form the initial
  guesses for $U_{m n \mathbf{k}}$.
We underline that while the gauge of Bloch wavefunctions $\ket{\psi_{m
        \mathbf{k}}}$ is arbitrary, \cref{eq:wan_init_proj} is invariant to such gauge
  freedom: suppose $\ket{\psi_{i \mathbf{k}}^{\prime} }$ are also solutions of
  the electronic structure problem, then $\ket{\psi_{i \mathbf{k}}^{\prime} }$
  are related to $\ket{\psi_{m \mathbf{k}} }$ by some unitary matrices
  $\ket{\psi_{i \mathbf{k}}^{\prime} } = \sum_m \ket{\psi_{m \mathbf{k}} } U_{m i
        \mathbf{k}}$; thus $\ket{ \phi_{n \mathbf{k}} } = \sum_m \ket{\psi_{m
        \mathbf{k}}} \braket{\psi_{m \mathbf{k}} | g_{n} } = \sum_m \sum_i
    \ket{\psi^{\prime}_{i \mathbf{k}}} U_{i m \mathbf{k}}^* U_{m i \mathbf{k}}
    \braket{\psi^{\prime}_{i \mathbf{k}} | g_{n} } = \sum_i \ket{\psi^{\prime}_{i
    \mathbf{k}}} \braket{\psi^{\prime}_{i \mathbf{k}} | g_{n} }$ does not depend on
  the gauge of Bloch wavefunctions, where superscript $*$ denotes conjugate
  transpose.
For entangled bands, the ``standard'' disentanglement approach \cite{Souza2001}
  uses energy windows to choose the disentanglement and frozen manifolds:
  \begin{inparaenum}[(a)] \item an (outer) disentanglement window that includes a
  large set of Bloch states, which can be mixed together to obtain a smaller
  disentangled manifold; \item an (inner) frozen window that specifies a smaller
  set of Bloch states (often states around Fermi energy) which are kept unchanged
  in the final disentangled manifold.
\end{inparaenum}

Since in the following sections the present results are compared with SCDM, we
  also summarize the SCDM procedure here.
The SCDM method \cite{Damle2015} starts from the real-space density matrix
  \densitymatrix where $P_{\mathbf{k}} = \sum_{m=1}^{J_{\mathbf{k}}} \ket{\psi_{m
        \mathbf{k}}} \bra{\psi_{m \mathbf{k}}}$, and uses QR factorization with column
  pivoting (QRCP) to decompose \densitymatrix into a set of localized real-space
  orbitals, thanks to the near-sightedness principle \cite{Prodan2005,Benzi2013}
  stating that the matrix elements \densitymatrix decay exponentially with the
  distance between two points $\mathbf{r}$ and $\mathbf{r}^{\prime}$ in
  insulating systems.
While storing the full \densitymatrix is memory intensive (it has size
  $N_{\mathbf{r}} \times N_{\mathbf{r}}$, where $N_{\mathbf{r}}$ is the number of
  real-space grid points), one can equivalently decompose the matrix formed by
  the real-space Bloch wavefunctions $\Psi_{\mathbf{k}}^* = [\psi_{1 \mathbf{k}},
    \dots, \psi_{J_{\mathbf{k}} \mathbf{k}}]^*$, which has a smaller size
  $J_{\mathbf{k}} \times N_{\mathbf{r}}$.
For periodic systems, often the choice of columns in the QRCP algorithm can be
  performed using the wavefunctions at the $\Gamma$ point only ($\Psi_{\Gamma}$)
  \cite{Damle2017}, and the same column selection is then used for all other
  \kpts.
For entangled bands, since the density matrix is not continuous across the
  \kpts, one can construct a quasi-density matrix (or equivalently a matrix of
  wavefunctions) $\sum_{m=1}^{J_{\mathbf{k}}} \ket{\psi_{m \mathbf{k}}}
    f(\varepsilon_{m \mathbf{k}}) \bra{\psi_{m \mathbf{k}}}$, where
  $f(\varepsilon_{m \mathbf{k}})$ is a smooth function of the energy eigenvalues
  $\varepsilon_{m \mathbf{k}}$, specifying the target energy window for the
  constructed MLWFs.
Often the complementary error function $\frac{1}{2} \erfc(\frac{\varepsilon -
      \mu}{\sigma})$ is chosen as $f(\varepsilon)$, and the choice of $\mu$ and
  $\sigma$ determines the shape of MLWFs, as well as band-interpolation quality.
Using projectability, defined later in \cref{eq:projectability_all}, $\mu$ and
  $\sigma$ can be automatically chosen, thus automating the Wannierization
  process \cite{Vitale2020}.

\section{Results and Discussions}
\subsection{Pseudo-atomic-orbital projections}
In addition to the hydrogenic orbitals discussed above, alternative starting
  guesses for the Wannierization can be used.
For instance, in pseudopotential plane-wave methods, PAOs are localized
  orbitals originating from the pseudopotential generation procedure
  \cite{Agapito2016}.
In this procedure, for each element, atomic wavefunctions of an isolated atom
  are pseudized to remove the radial nodes and are localized functions around the
  atom; spherical harmonics with well-defined angular-momentum character ($s$,
  $p$, $d$, or $f$) are chosen for their angular dependency.
Then, the PAOs are summed over lattice points with appropriate phases to obtain
  Bloch sums, Fourier transformed to a plane-wave basis, \lowdin-orthonormalized,
  and finally taken as the projectors for initial projections.
PAOs are commonly used for analyzing the orbital contributions to band
  structures, as the basis set for non-iterative construction of TB Hamiltonians
  \cite{Agapito2016}, or as projectors in DFT+Hubbard calculations
  \cite{PhysRevMaterials.5.104402}.

\begin{figure}[tb]
  \includegraphics[width=\linewidth,max height=0.6\textheight]{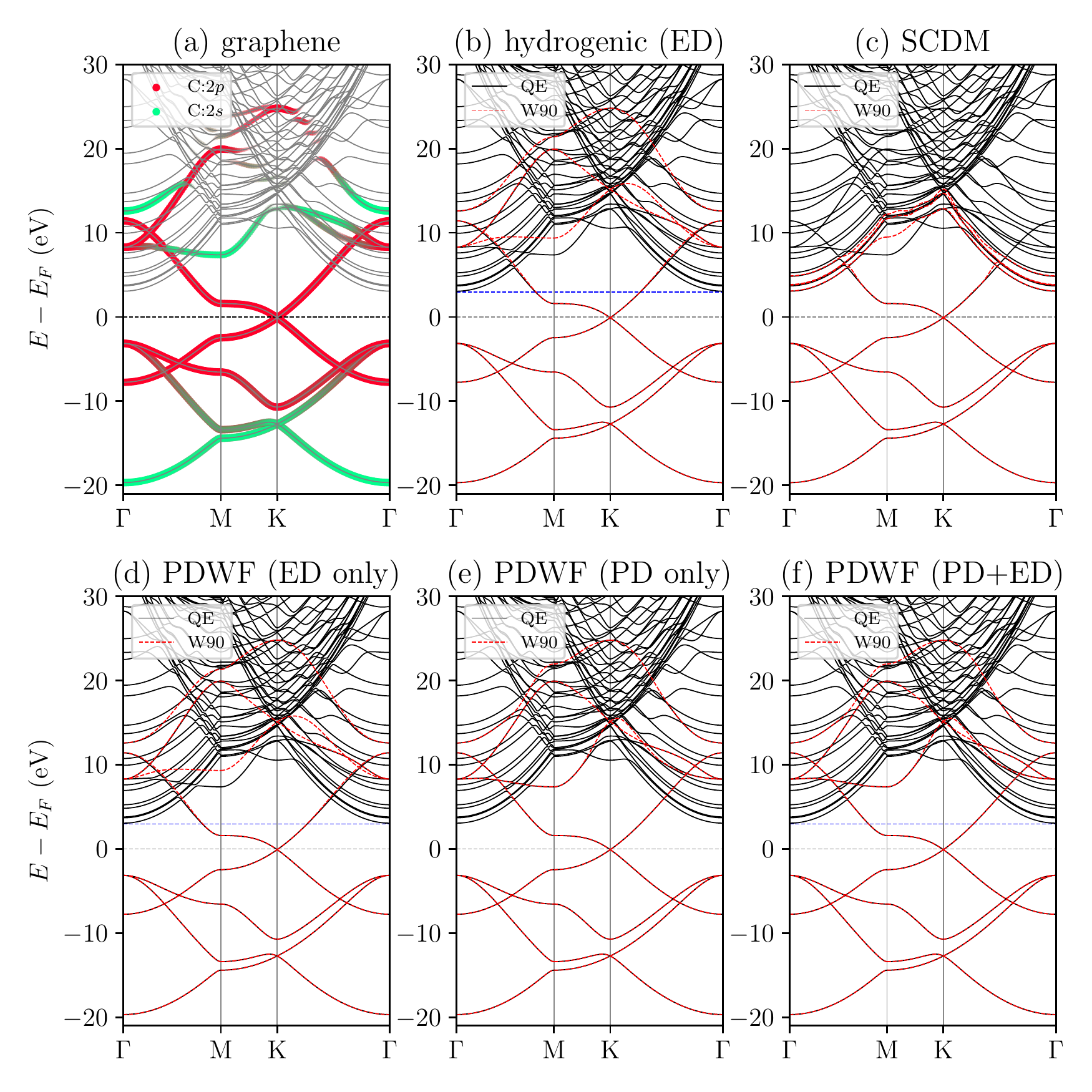}
  \begin{subcaptiongroup}
    \phantomsubcaption\label{fig:bands_c2&projband}
    \phantomsubcaption\label{fig:bands_c2&aao_wd}
    \phantomsubcaption\label{fig:bands_c2&scdm}
    \phantomsubcaption\label{fig:bands_c2&nao_wd}
    \phantomsubcaption\label{fig:bands_c2&nao_pd}
    \phantomsubcaption\label{fig:bands_c2&nao_pwd}
  \end{subcaptiongroup}
  \cprotect\caption{\textbf{Comparisons of graphene band structures interpolated using different methods.}
    \subref{fig:bands_c2&projband} DFT band structure, shown as grey lines.
    The colored dots represent the projectabilities onto carbon $2s$ (green) and
      $2p$ (red) orbitals.
    The size of each dot is proportional to the total projectability
      $p_{m\mathbf{k}}$ of the band $m$ at \kpt $\mathbf{k}$; see
      \cref{eq:projectability_all}.
    For a detailed plot of total projectability, see
      \cref{supp-fig:projband_graphene}.
    Comparisons of the original and the Wannier-interpolated bands for
      \subref{fig:bands_c2&aao_wd} hydrogenic projections with energy disentanglement
      (ED), \subref{fig:bands_c2&scdm} SCDM, \subref{fig:bands_c2&nao_wd} PAO with
      ED, \subref{fig:bands_c2&nao_pd} PAO with projectability disentanglement (PD),
      and \subref{fig:bands_c2&nao_pwd} PAO with PD+ED.
    The Fermi energy \ef (horizontal black dashed line) is at zero; the horizontal
      blue dashed line denotes the top of the inner energy window, where applicable.
  } \label{fig:bands_c2}
\end{figure}

In order to understand the contribution of each orbital $\ket{ g_{n} }$ to a
  Bloch state $\ket{ \psi_{m \mathbf{k}} }$, we define a measure of
  projectability as the square of the inner product between $\ket{ \psi_{m
        \mathbf{k}} }$ and $\ket{ g_{n} }$:
  \begin{equation}
    \label{eq:projectability}
    p_{nm\mathbf{k}} = \left| \braket{ g_{n} | \psi_{m \mathbf{k}} } \right|^2;
  \end{equation}
  the projectability of $\ket{ \psi_{m \mathbf{k}} }$ onto all
  PAOs is then defined as \begin{equation} \label{eq:projectability_all}
  p_{m\mathbf{k}} = \sum_{n} p_{nm\mathbf{k}}.
\end{equation}
If the projectors $\ket{ g_{n} }$ are complete for $\ket{ \psi_{m \mathbf{k}}
    }$, then $p_{m\mathbf{k}} = 1$.
The band projectability is a very useful criterion to identify the orbital
  character of the bands; this is exemplified in \cref{fig:bands_c2&projband},
  where we show the projectability of the bands of graphene onto $2s$ and $2p$
  PAOs for carbon.
It is immediately apparent how one can easily identify states in the conduction
  manifold that have a strong $2p$ and $2s$ component.

Compared with the hydrogenic projections, which is the method used by default
  in \WAN \cite{Pizzi2020} and its interface code to \QE \cite{Giannozzi2020a}
  (called \PtoW), PAOs are better adapted to each element since they come exactly
  from the pseudopotential used in the actual solid-state calculation.
Moreover, in pseudopotentials with semicore states, the PAOs for semicores are
  nodeless and those for valence wavefunctions have at least one radial node (so
  as to be orthogonal to the semicore states with same angular momentum); thus
  band projectability can clearly differentiate semicore from valence, making
  PAOs more convenient than the hydrogenic orbitals, for which the user would
  need to manually set the correct radial functions for both semicore and valence
  projectors.
For these reasons, we use in this work the PAOs as initial and more accurate
  projections.
If needed, higher energy orbitals not included in the pseudopotential file can
  be constructed, for example, using solutions of Schr\"odinger equation under
  confinement potential \cite{Ozaki2003,Ozaki2004} (see also discussion in
  \cref{sec:add_paos}).

\subsection{Projectability disentanglement}\label{sec:proj_dis}
As mentioned, the standard disentanglement approach selects the disentanglement
  and frozen manifolds via two energy windows \cite{Souza2001}.
We refer to this as energy disentanglement (ED).
However, since bands have dispersions across the BZ, a fixed window for all
  \kpts might not be an optimal choice.
Taking the graphene band structure (\cref{fig:bands_c2&projband}) as an
  example, the bands with large projectability are mixed with many free-electron
  bands with zero projectability (grey bands in the conduction region).
In this case, one is faced with several options for the outer and inner energy
  windows, each with different shortcomings: \begin{inparaenum}[(a)] \item If the
  inner window includes free-electron bands, the final MLWFs are mixtures of
  $2s$, $2p$ atomic orbitals and free-electron bands, delocalizing the resulting
  MLWFs; \item if the outer window excludes both the free-electron bands and the
  atomic-orbital states inside free-electron bands, the WFs lack the anti-bonding
  part of the bonding/anti-bonding closure \cite{Thygesen2005}, again degrading
  the localization of WF; \item if the upper bound of the inner window is set to
  its maximal allowed value, \ie the blue dashed line positioned at the minimum
  of free-electron bands in \cref{fig:bands_c2&aao_wd}, and all the DFT
  eigenstates are included in the outer window, the disentanglement algorithm
  \cite{Souza2001} will extract an optimally smooth manifold, at the expense of
  decreasing the chemical representability of the atomic-orbital bands in the
  free-electron region; in other words, the MLWFs obtained lose the information
  of the TB atomic orbitals in this chemical environment (see
  \cref{fig:bands_c2&aao_wd}).
\end{inparaenum}

The graphene case highlights the limitations of the standard ED.
Instead, we propose here to select the disentanglement and frozen manifolds
  based on the projectability $p_{m\mathbf{k}}$ of each state on the chosen PAOs
  (\ie, states are selected irrespective of their energy, but rather based on
  their chemical representativeness).
Specifically, we select states based on two thresholds \pmin and \pmax:
  \begin{inparaenum}[(a)] \item If $p_{m\mathbf{k}} <$ \pmin, the state
  $\psi_{m\mathbf{k}}$ is discarded.
\item If $p_{m\mathbf{k}} \ge$ \pmax, the state $\psi_{m\mathbf{k}}$ is kept
identically.
\end{inparaenum}
Crucially, all states for which \pmin $\le p_{m\mathbf{k}} <$ \pmax are thrown
  in the disentanglement algorithm.
Optimal numerical values for \pmin and \pmax are discussed later.
In the case of graphene, \pmax identifies the fully atomic-orbital states
  inside the free-electron bands, while \pmin removes the fully free-electron
  bands from the disentanglement process, preventing the mixing of atomic and
  free-electron states.
The two thresholds \pmin and \pmax constitute the parameters of the
  disentanglement process, replacing the four defining energy windows (the lower
  and upper bounds of the outer and inner energy windows).
We note that projectability disentanglement is different from partly-occupied
  WF \cite{Thygesen2005,Thygesen2005a} in that the latter uses an energy window
  to select frozen states and minimizes the total spread functional directly,
  while projectability disentanglement selects the localized states using
  projectability instead of a constant energy window across \kpts.
In fact, one can combine projectability disentanglement with a variational
  formulation \cite{Damle2019} to construct MLWFs by minimizing directly the
  total spread functional.

Ideally, if PAOs were always a complete set to describe valence and
  near-Fermi-energy conduction bands, the PD would select the most relevant Bloch
  states and accurately interpolate these DFT bands.
However, since the PAOs are fixed orbitals from isolated single-atom
  calculations for each element, if the chemical environment in the crystal
  structure is significantly different from that of pseudopotential generation,
  then the total projectability $p_{m\mathbf{k}}$ might be smaller than $1$ for
  bands around the conduction band minimum (CBM) or even for valence bands.
In such cases, one solution is to increase the number of PAOs, \ie, adding more
  projectors with higher angular momentum, as we will discuss in
  \cref{sec:add_paos}.
However, since one almost always wants to correctly reproduce valence bands
  (plus possibly the bottom of the conduction) but at the same time keep the
  Wannier Hamiltonian small for computational reasons, we suggest to additionally
  freeze all the states that sit below the Fermi energy in metals (or below the
  CBM for insulators) and also those a few \si{eV} above (typically, \SI{2}{eV}
  or so).
Such a combination of PD+ED gives accurate interpolation of bands below and
  around the Fermi energy (or band edges for insulators), as well as maximally
  restoring the atomic-orbital picture.

We stress here that, even if we call the resulting Wannier functions PDWFs for
  clarity, our optimal suggestion is to always also freeze the states in the
  energy window mentioned above, as we discuss in the next sections.

\subsection{Comparison}
We choose four prototypical materials to discuss the present method: graphene,
  silicon, copper, and strontium vanadate (\SrVOthree).
Graphene is a difficult case where atomic-orbital states highly mix with
  free-electron bands; silicon tests the Wannierization of both valence and
  conduction bands of an insulator; copper is a test on a metal; and \SrVOthree
  represents the class of (metallic) perovskites.
We compare the shapes, centers, and spreads of the resulting MLWFs using the
  five methods mentioned earlier: hydrogenic projection with ED (\ie, the
  standard approach), SCDM, PAO projection with ED, PAO projection with PD, and
  PAO projection with PD+ED.

\subsubsection{Graphene}
\begin{figure}[tb]
  \includegraphics[width=\linewidth,max height=0.6\textheight]{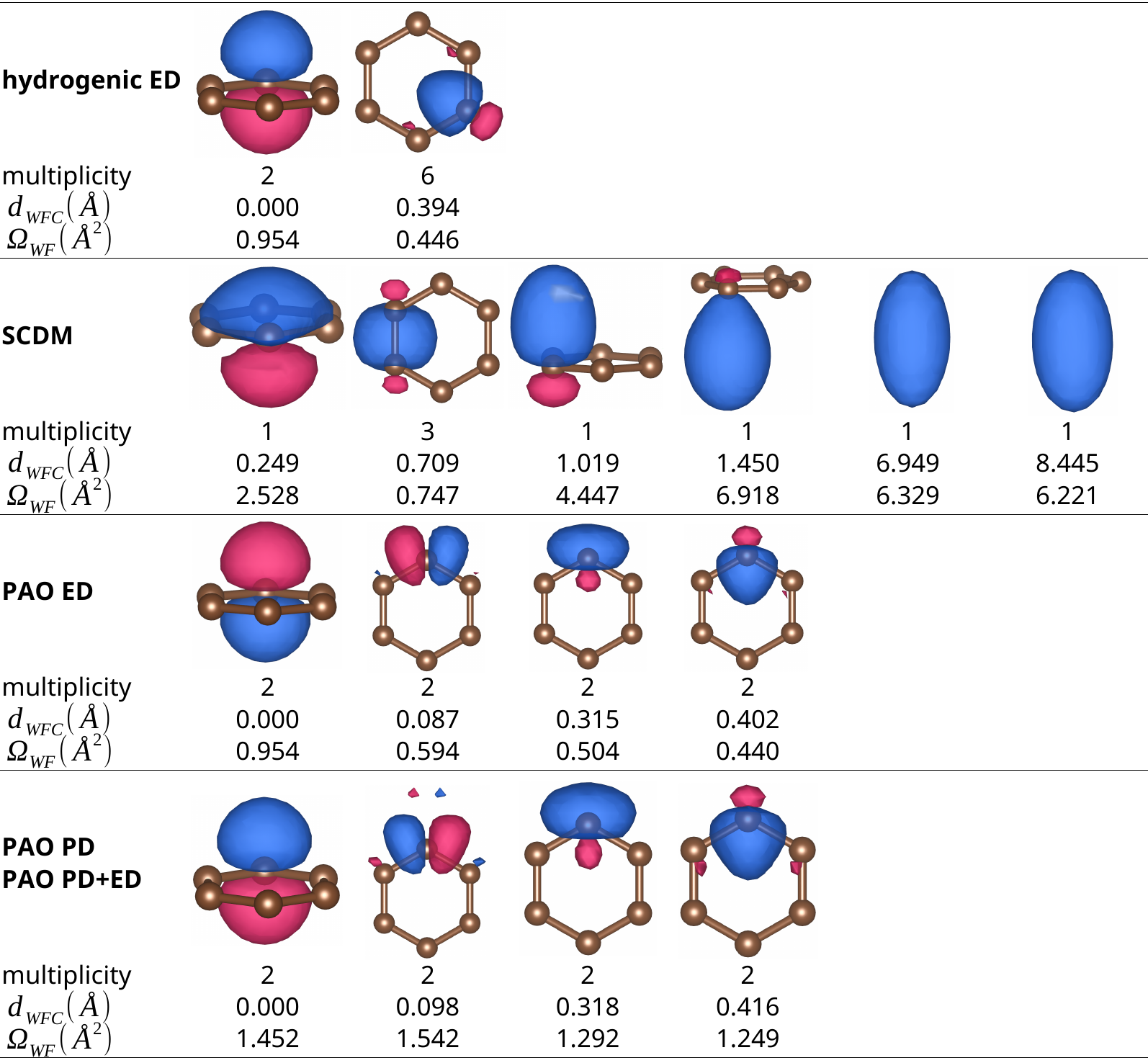}
  \cprotect\caption{\textbf{Graphene MLWFs: shapes, centers, and spreads obtained using different methods.}
    \dwfc is the distance of the WF center from
    the nearest-neighbor atom,
    and \omegawf is the MLWF spread.
    The multiplicity is the number of equivalent MLWFs, \ie having the same \dwfc,
      \omegawf, and shape, but different orientations.
  }
  \label{fig:wf_shape_graphene}
\end{figure}

The original and interpolated band structures for the five methods discussed
  are shown in
  \cref{fig:bands_c2&aao_wd,fig:bands_c2&scdm,fig:bands_c2&nao_wd,fig:bands_c2&nao_pd,fig:bands_c2&nao_pwd}.
The blue dashed lines in
  \cref{fig:bands_c2&aao_wd,fig:bands_c2&nao_wd,fig:bands_c2&nao_pwd} indicate
  the top of the inner energy window, which is set optimally (and manually) to
  just below the free-electron bands, to freeze as much as possible the
  atomic-orbital states but exclude any free-electron state.
For PD and PD+ED, we choose \pmax = 0.85 and \pmin = 0.02 (we will discuss
  later on the choice of these thresholds).
Comparing \cref{fig:bands_c2&nao_wd} and \cref{fig:bands_c2&aao_wd}, one sees
  that ED produces similar bands irrespective of using hydrogenic or PAO
  projection.
However, as shown in \cref{fig:wf_shape_graphene} (first and third row), the
  MLWFs for the two cases fall into slightly different minima: MLWFs from
  hydrogenic projection with ED are $p_z$ and hybridized $s \pm p$ orbitals
  pointing towards the center of the hexagon, while MLWFs from PAO with ED are
  $p_z$, $p_x$, and $s \pm p_y$.
This is due to the fact that the PAO projections guide the minimization towards
  spherical harmonics, while the hydrogenic projections are farther away from
  such local minimum and the optimization algorithm happens to escape and
  converge to a better minimum.
A possible future work is to introduce more advanced optimization algorithms to
  improve the convergence of maximal localization.
Both the PAO with PD and PAO with PD+ED cases reach the same set of MLWFs,
  $p_z$, $p_x$, and $s \pm p_y$, but with larger spreads than the PAO with ED,
  since the PD and PD+ED freeze more states, giving thus less freedom for maximal
  localization.
Nevertheless, the interpolated bands of the PAO with PD and PAO with PD+ED
  cases can much better reproduce the atomic-orbital states inside the
  free-electron bands.
Finally, compared to other cases, SCDM includes some free-electron bands, some
  of which can be even reproduced by the Wannier interpolation.
However, in order to follow those free-electron bands, abrupt changes of
  character and band derivative are needed in the conduction band.
As required by Nyquist--Shannon sampling theorem\cite{Oppenheim1997}, this
  results in a denser $\mathbf{k}$-space sampling needed to obtain a good
  interpolation quality.
Moreover, the MLWFs are much more delocalized and do not resemble atomic
  orbitals: as shown in \cref{fig:wf_shape_graphene}, the last two MLWFs for SCDM
  are floating away from the graphene 2D lattice, blurring the TB picture of
  atomic orbitals in solids.

\subsubsection{Silicon}\label{sec:silicon}
\begin{figure}[tb]
  \includegraphics[width=\linewidth,max height=0.6\textheight]{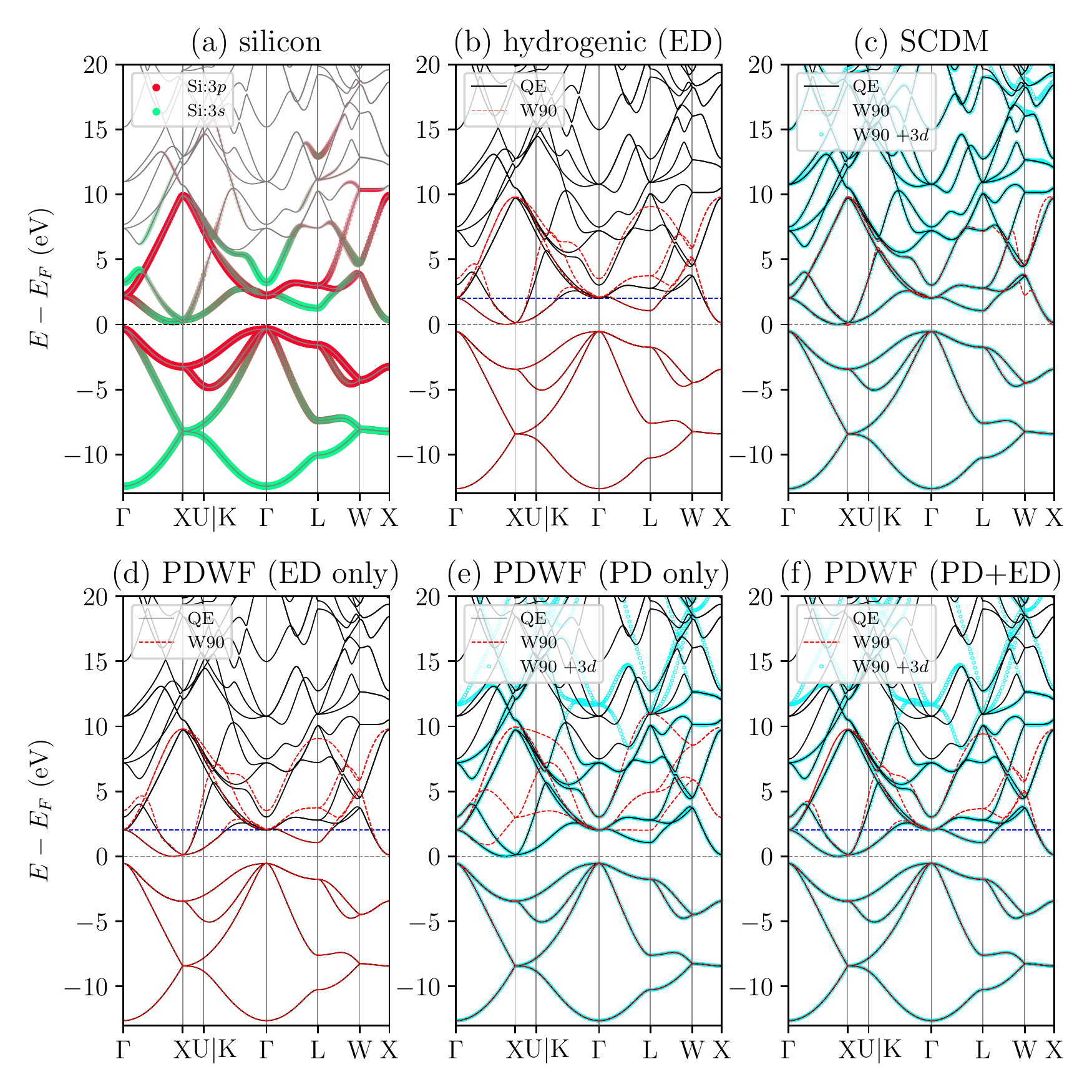}
  \begin{subcaptiongroup}
    \phantomsubcaption\label{fig:bands_si2&projband}
    \phantomsubcaption\label{fig:bands_si2&aao_wd}
    \phantomsubcaption\label{fig:bands_si2&scdm}
    \phantomsubcaption\label{fig:bands_si2&nao_wd}
    \phantomsubcaption\label{fig:bands_si2&nao_pd}
    \phantomsubcaption\label{fig:bands_si2&nao_pwd}
  \end{subcaptiongroup}
  \cprotect\caption{\textbf{Comparisons of silicon band structures interpolated
      using different methods.}
    \subref{fig:bands_si2&projband} DFT band structure, shown as grey lines.
    The colored dots represent the projectabilities of silicon $3s$ (green) and
      $3p$ (red) orbitals.
    The size of the dot is proportional to the total projectability
      $p_{m\mathbf{k}}$ of the band $m$ at \kpt $\mathbf{k}$.
    For a detailed plot of total projectability, see
      \cref{supp-fig:projband_silicon}.
    Comparisons of the original and the Wannier-interpolated bands for
      \subref{fig:bands_si2&aao_wd} hydrogenic projections with ED,
      \subref{fig:bands_si2&scdm} SCDM, \subref{fig:bands_si2&nao_wd} PAO with ED,
      \subref{fig:bands_si2&nao_pd} PAO with PD, and \subref{fig:bands_si2&nao_pwd}
      PAO with PD+ED.
    The CBM (horizontal black dashed line) is at zero; the horizontal blue dashed
      line denotes the top of the inner energy window, \ie, CBM + \SI{2}{eV}, where
      applicable.
    Note in \subref{fig:bands_si2&scdm}, \subref{fig:bands_si2&nao_pd}, and
      \subref{fig:bands_si2&nao_pwd}, the cyan lines with circle markers show the
      interpolated bands obtained including also 3d orbitals, and consequently
      increasing the dimensionality of the disentangled manifold.
    These additional states are beneficial because of the presence of an intrinsic
      $d$ component at the bottom of the conduction manifold, and lead to more
      accurate band interpolations.
  } \label{fig:bands_si2}
\end{figure}

The SCDM method obtains four front-bonding and four back-bonding MLWFs, while
  all other cases lead to atom-centered $s$ and $p$ MLWFs, as shown in
  \cref{supp-fig:wf_shape_silicon}.
While overall the SCDM bands (\cref{fig:bands_si2&scdm}) seem to reproduce
  relatively better the higher conduction bands, they fail to correctly reproduce
  the bottom of the conduction band near the X point, induce more wiggles around
  X and W, and have much larger spreads.
Due to the low projectability of Bloch states around X ($p_{m\mathbf{k}}$
  around 0.83), the CBM is not correctly reproduced in the PAO with PD, as these
  are not frozen in PD with the current choice of \pmax = 0.95 and \pmin = 0.01.
To explicitly freeze the CBM, \pmax would need to be lowered below 0.83.
However, such kind of decrease will also result in freezing some high-energy
  conduction bands, degrading the localization.
PD+ED overcomes this by explicitly freezing the near-Fermi-energy and
  low-projectability states at the CBM, but still only freezing those
  atomic-orbital states in the high-energy conduction bands that possess high
  projectability (see \cref{fig:bands_si2&nao_pwd}), thus improving band
  interpolation.
We note that the lower projectability of silicon CBM is intrinsic to the
  material---its CBM also includes $3d$ character.
Therefore, by adding $d$ PAOs, the CBM projectability increases (from 0.83 to
  0.99) and one can restore a high-quality band-structure interpolation within
  the PD method: as shown in \cref{fig:bands_si2&nao_pd}, the low-energy
  conduction bands are correctly reproduced once we regenerate a silicon
  pseudopotential including $3d$ PAOs.
Therefore, PD is sufficient to obtain an accurate band interpolation if enough
  PAOs are included (we will also discuss this later in \cref{sec:add_paos}).
For completeness, we show the SCDM interpolation using the regenerated
  pseudopotential in \cref{fig:bands_si2&scdm}: the added $d$ PAOs help select a
  larger manifold thanks to the increased projectability, enabling SCDM to
  reproduce higher conduction bands, as well as fixing the wrong interpolation at
  the W point.
Moreover, additional PAOs can also benefit ED, since the frozen window can be
  enlarged to reproduce more states.
In general, adding more PAOs improves interpolation quality in cases where the
  target bands have low projectability, at the price of increased computational
  cost.
PD+ED is a better option for reaching a good interpolation accuracy while
  keeping the size of the corresponding TB model small.

\subsubsection{Copper and \SrVOthree}
Results for copper and \SrVOthree are only shown in the SI
  (\cref{supp-fig:bands_cu,supp-fig:wf_shape_cu,supp-fig:bands_srvo3,supp-fig:wf_shape_srvo3}),
  since the conclusions are the same: PD+ED consistently provides the best
  interpolation quality among all methods we consider, while not requiring to
  increase the size of the Hamiltonian model, and results in WFs that resemble
  atomic orbitals or their hybridization.

\subsection{High-throughput verification}
In this section we discuss the applicability of the present PDWF method to
  obtain, in a fully automated way and without user input, WFs for any material.
In order to assess quantitatively its performance, we compare it to SCDM, that
  can also be fully automated (see Ref.~\cite{Vitale2020}).

In all results that follow, we exclude semicore orbitals in both methods, since
  these low-energy states correspond to almost flat bands and do not play any
  role in the chemistry of the materials.
We compare quantitatively the band interpolation quality between the two
  methods and the corresponding WF centers and spreads on the 200-structure set
  used in Ref.~\cite{Vitale2020} for both occupied and unoccupied bands,
  totalling \num{6818} MLWFs for each method.
In accordance with Refs.~\cite{Vitale2020,Prandini2018}, the band interpolation
  quality is measured by the average band distance,
  \begin{equation}
    \label{eq:bandsdist} \eta_{\nu} = \sqrt{\frac {\sum_{n\mathbf{k}}
    \tilde{f}_{n\mathbf{k}} (\epsilon_{n\mathbf{k}}^{\text{DFT}} -
    \epsilon_{n\mathbf{k}}^{\text{Wan}})^2 } {\sum_{n\mathbf{k}}
    \tilde{f}_{n\mathbf{k}}} },
  \end{equation}
  and the max band distance,
  \begin{equation}
    \label{eq:bandsdist_max} \eta_{\nu}^{\max} =
    \max_{n\mathbf{k}} \left( \tilde{f}_{n\mathbf{k}} \left|
    \epsilon_{n\mathbf{k}}^{\text{DFT}} - \epsilon_{n\mathbf{k}}^{\text{Wan}}
    \right| \right),
  \end{equation}
  where $\tilde{f}_{n\mathbf{k}} =
    \sqrt{f^{\text{DFT}}_{n\mathbf{k}}(E_{F}+\nu, \sigma)
      f^{\text{Wan}}_{n\mathbf{k}}(E_{F}+\nu, \sigma)}$ and $f(E_{F}+\nu, \sigma)$ is
  the Fermi-Dirac distribution.
Here $E_{F}+\nu$ and $\sigma$ are fictitious Fermi levels and smearing widths
  which we choose for comparing a specific range of bands.
Since the Wannier TB model describes the low-energy valence electrons, it is
  expected that the band interpolation deviates from the original in the higher
  conduction band region.
Therefore, the higher $\nu$ is, the larger $\eta_{\nu}$ is expected to be.
In the following paragraphs, we will use $\eta_{0}$ and $\eta_{2}$ to compare
  bands below \ef and \ef + \SI{2}{eV}, respectively; $\sigma$ is always fixed at
  \SI{0.1}{eV}.

In the supplementary information \cref{supp-sec:200pages}, we provide
  comparisons between the Wannier-interpolated bands and the DFT bands for both
  PDWF and SCDM, their respective band distances, and the Hamiltonian decay plots
  for each of the 200 materials.
We discuss these properties in the following.

\subsubsection{Projectability thresholds and automation}
\begin{figure}[tb]
  \includegraphics[width=\linewidth,max height=0.6\textheight]{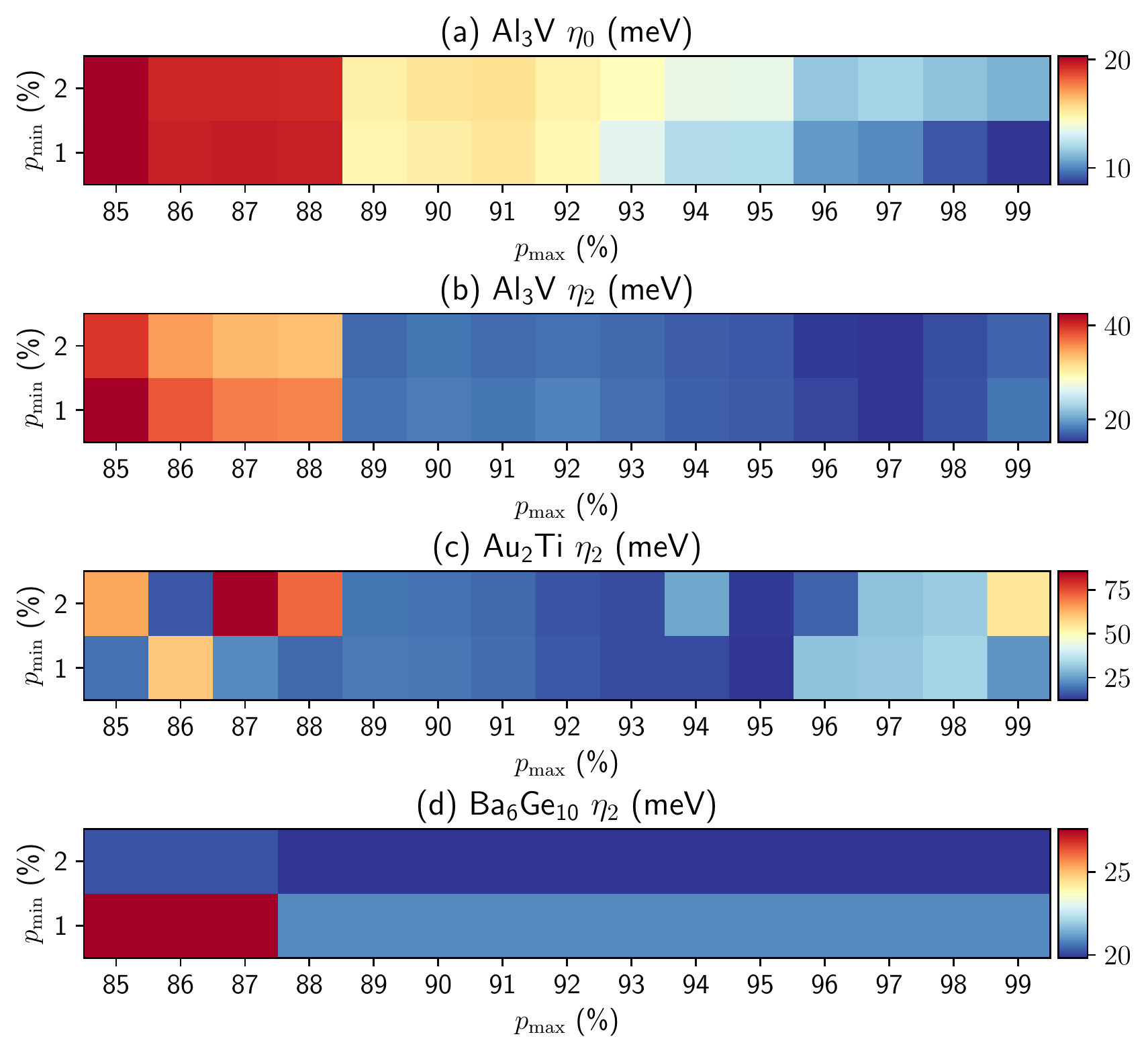}
  \begin{subcaptiongroup}
    \phantomsubcaption\label{fig:proj_thresholds&al3v_eta0}
    \phantomsubcaption\label{fig:proj_thresholds&al3v_eta2}
    \phantomsubcaption\label{fig:proj_thresholds&au2ti_eta2}
    \phantomsubcaption\label{fig:proj_thresholds&ba6ge10_eta2}
  \end{subcaptiongroup}
  \cprotect\caption{ \textbf{Quality of band interpolations: band
      distances for different choices of \pmin and \pmax.}
    \subref{fig:proj_thresholds&al3v_eta0} $\eta_0$ of Al$_3$V,
      \subref{fig:proj_thresholds&al3v_eta2} $\eta_2$ of Al$_3$V,
      \subref{fig:proj_thresholds&au2ti_eta2} $\eta_2$ of Au$_2$Ti, and
      \subref{fig:proj_thresholds&ba6ge10_eta2} $\eta_2$ of Ba$_6$Ge$_{10}$.
    Note the color scale is different for each plot.
  }
  \label{fig:proj_thresholds}
\end{figure}

For PDWF, we set the maximum of the inner window to the Fermi energy +
  \SI{2}{eV} for metals, or to the CBM + \SI{2}{eV} for insulators, to fully
  reproduce states around Fermi energy or the band edges.
We also specify the two additional parameters \pmin and \pmax.
From our tests, in most cases \pmax = 0.95 and \pmin = 0.01 already produce
  very good results.
However, since chemical environments vary across different crystal structures,
  the two parameters are not universal and influence the quality of band
  interpolation.
\Cref{fig:proj_thresholds} shows the variation of band distances \wrt \pmin and
  \pmax for several materials.
For \AlthreeV
  (\cref{fig:proj_thresholds&al3v_eta0,fig:proj_thresholds&al3v_eta2}), $\eta_0$
  and $\eta_2$ reach a minimum at two different sets of parameters, \ie, \pmax =
  0.99, \pmin = 0.01 and \pmax = 0.97, \pmin = 0.01, respectively.
In some cases, the variation of $\eta$ \wrt \pmax and \pmin can be
  non-monotonic and display multiple local minima: For instance, in \AutwoTi
  (\cref{fig:proj_thresholds&au2ti_eta2}) at \pmin = 0.01, $\eta_2$ decreases
  from \pmax = 0.90 to 0.95 but increases from \pmax = 0.95 to 0.98 and finally
  reaches a local minimum at \pmax = 0.99.
In other cases, $\eta$ can be quite stable and largely independent of the
  parameters: \eg, for \BasixGeten (\cref{fig:proj_thresholds&ba6ge10_eta2}),
  $\eta_2$ reaches the same minimum for \pmax = 0.99 to 0.88.

Therefore, we implement an iterative optimization workflow to automatically
  find the optimal values for \pmax and \pmin, in order to fully automate the
  Wannierization procedure.
The workflow is released as part of the \texttt{aiida-wannier90-workflows}
  package \cite{aiidaW90}.
First, we run a QE band structure workflow to get the reference DFT bands for
  calculating $\eta_2$; in addition, the DFT bands are also used to calculate the
  band gap of the material.
Second, we run an optimization workflow with the following settings: The
  maximum of the inner window is set to Fermi energy + \SI{2}{eV} for metals and
  CBM + \SI{2}{eV} for insulators, respectively; \pmax and \pmin are set to the
  defaults of 0.95 and 0.01, respectively.
Third, if the average band distance $\eta_2$ is less than a threshold (set to
  \SI{10}{meV} here), the workflow stops; otherwise, the workflow iterates on a
  mesh of \pmax and \pmin, \ie \pmax decreasing from 0.99 to 0.80 with step size
  -0.01, and \pmin = 0.01 or 0.02, until $\eta_2 \le$ threshold.
If $\eta_2$ is still larger than the threshold after exhausting all the
  parameter combinations, the workflow will output the minimum-$\eta_2$
  calculation.

\subsubsection{Band distance}
\begin{figure}[!htb]
  \includegraphics[width=\linewidth,max height=0.6\textheight]{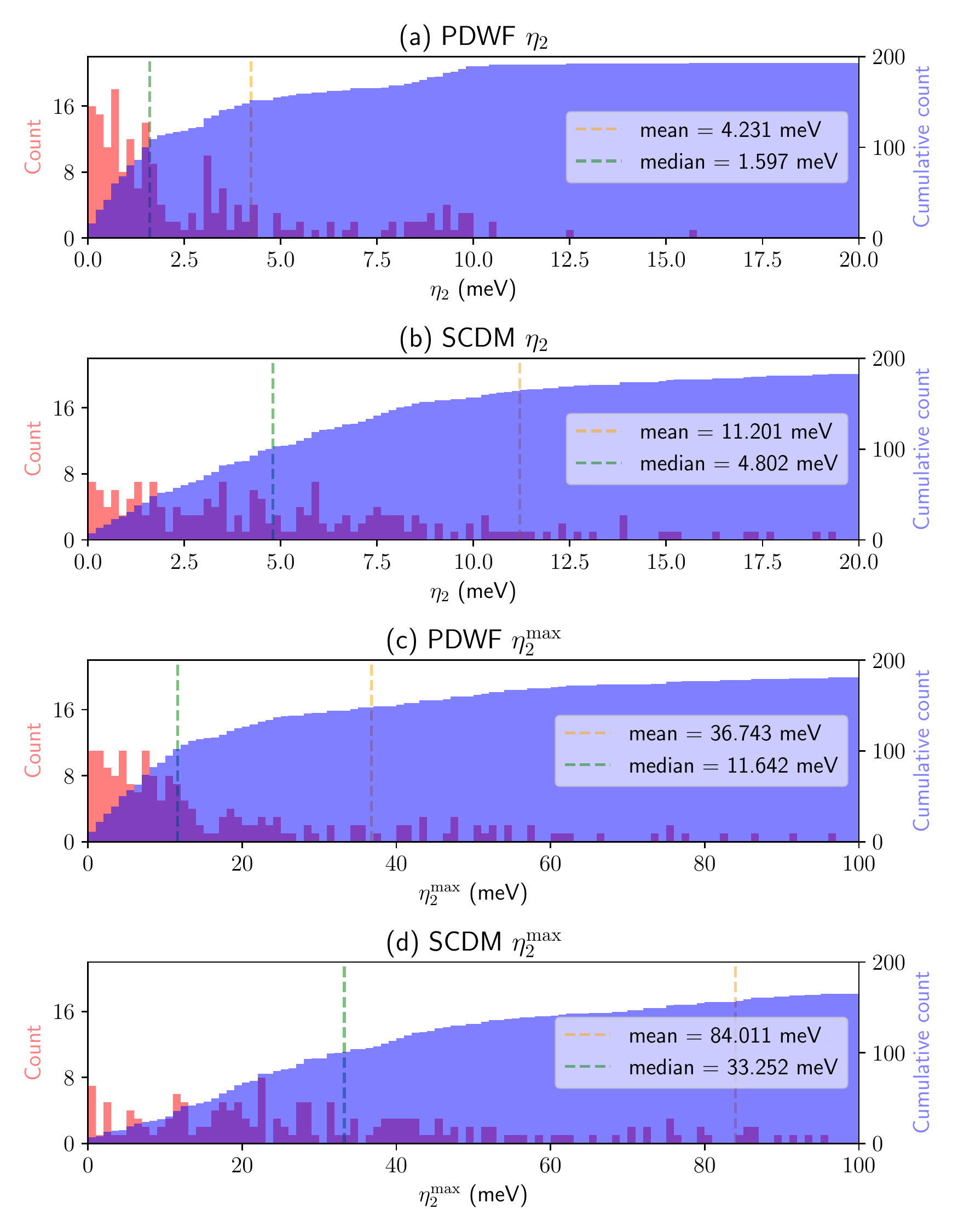}
  \begin{subcaptiongroup}
    \phantomsubcaption\label{fig:bandsdist_scdm200&eta_nao}
    \phantomsubcaption\label{fig:bandsdist_scdm200&eta_scdm}
    \phantomsubcaption\label{fig:bandsdist_scdm200&etamax_nao}
    \phantomsubcaption\label{fig:bandsdist_scdm200&etamax_scdm}
  \end{subcaptiongroup}
  \cprotect\caption{ \textbf{Histogram (red) and cumulative histogram (blue) of
      the band distances $\eta_2$ and $\eta_2^{\max}$ for 200 reference structures.}
    \subref{fig:bandsdist_scdm200&eta_nao} $\eta_2$ of PDWF,
      \subref{fig:bandsdist_scdm200&eta_scdm} $\eta_2$ of SCDM,
      \subref{fig:bandsdist_scdm200&etamax_nao} $\eta_2^{\max}$ of PDWF, and
      \subref{fig:bandsdist_scdm200&etamax_scdm} $\eta_2^{\max}$ of SCDM.
    The orange (green) vertical line is the mean (median) of the band distance for
      the 200 structures; their values are shown in the right of each panel; PDWF
      provides approximately an improvement by a factor of 3.
  }
  \label{fig:bandsdist_scdm200}
\end{figure}

To compare quantitatively the band interpolation quality of SCDM and PDWF, we
  Wannierize the 200 structures mentioned earlier and calculate their band
  distances with respect to the corresponding DFT bands.
We choose $\eta_2$ and $\eta_2^{\max}$ to compare near-Fermi-energy bands.
The histograms of the band distances for the 200 structures are shown in
  \cref{fig:bandsdist_scdm200}.
To directly compare SCDM and PDWF, the mean and median value of $\eta$ of the
  200 calculations are shown as vertical lines in each panel.
For PDWF, the mean $\eta_2$ is \SI{4.231}{meV}, to be compared with
  \SI{11.201}{meV} for SCDM.
For $\eta_2^{\max}$ (that is a more stringent test of the quality of
  interpolation) the PDWF method also performs better, with a $\eta_2^{\max} =$
  \SI{36.743}{meV} vs.
\SI{84.011}{meV} for SCDM.
We can also observe this trend in \cref{fig:bandsdist_scdm200}: For $\eta_2$
  and $\eta_2^{\max}$, the PDWF histogram bins are much more clustered towards
  $\eta = 0$.
Note that in the cumulative histograms of $\eta_2$, at $\eta$ = \SI{20}{meV},
  the PDWF cumulative count is closer to the total number of calculations (200).
This indicates that the PDWF has a higher success rate in reducing the
  interpolation error below \SI{20}{meV}.
Similarly, for $\eta_2^{\max}$, PDWF has a higher success rate in reducing the
  interpolation error under \SI{100}{meV} (to get a better overview of $\eta$ and
  $\eta^{\max}$, we further show the same histograms of $\eta$ in a wider range
  \SIrange{0}{100}{meV}, and $\eta^{\max}$ in range \SIrange{0}{500}{meV}, in
  \cref{supp-fig:bandsdist_scdm200_eta_range100,supp-fig:bandsdist_scdm200_etamax_range500}).
To reduce the effect of major outliers, we can also compare the interpolation
  accuracy of successful calculations, \ie, excluding the outlier calculations
  which have significantly large band distances.
As shown in \cref{supp-tab:bandsdist_additional}, the $\eta_2^{\le 20}$, \ie,
  the average of all the calculations for which $\eta_2 \le$ \SI{20}{meV},
  indicates that PDWF (\SI{2.922}{meV}) is twice as good as SCDM
  (\SI{5.280}{meV}), and also has a higher success rate: for $\eta_2^{\le 20}$,
  $193 / 200 = 96.5\%$ of the structures have $\eta_2 \le$ \SI{20}{meV}, while
  for SCDM it is $183 / 200 = 91.5\%$.
More details are listed in \cref{supp-tab:bandsdist_additional}.

In summary, PDWF provides more accurate and robust interpolations, especially
  for bands around the Fermi energy or the band gap edges, which are the most
  relevant bands for many applications.
Last but not least, a higher energy range can be accurately interpolated by
  increasing the number of PAOs (see \cref{sec:add_paos}).

\subsubsection{MLWF centers}
\begin{figure}[tb]
  \includegraphics[width=\linewidth,max height=0.6\textheight]{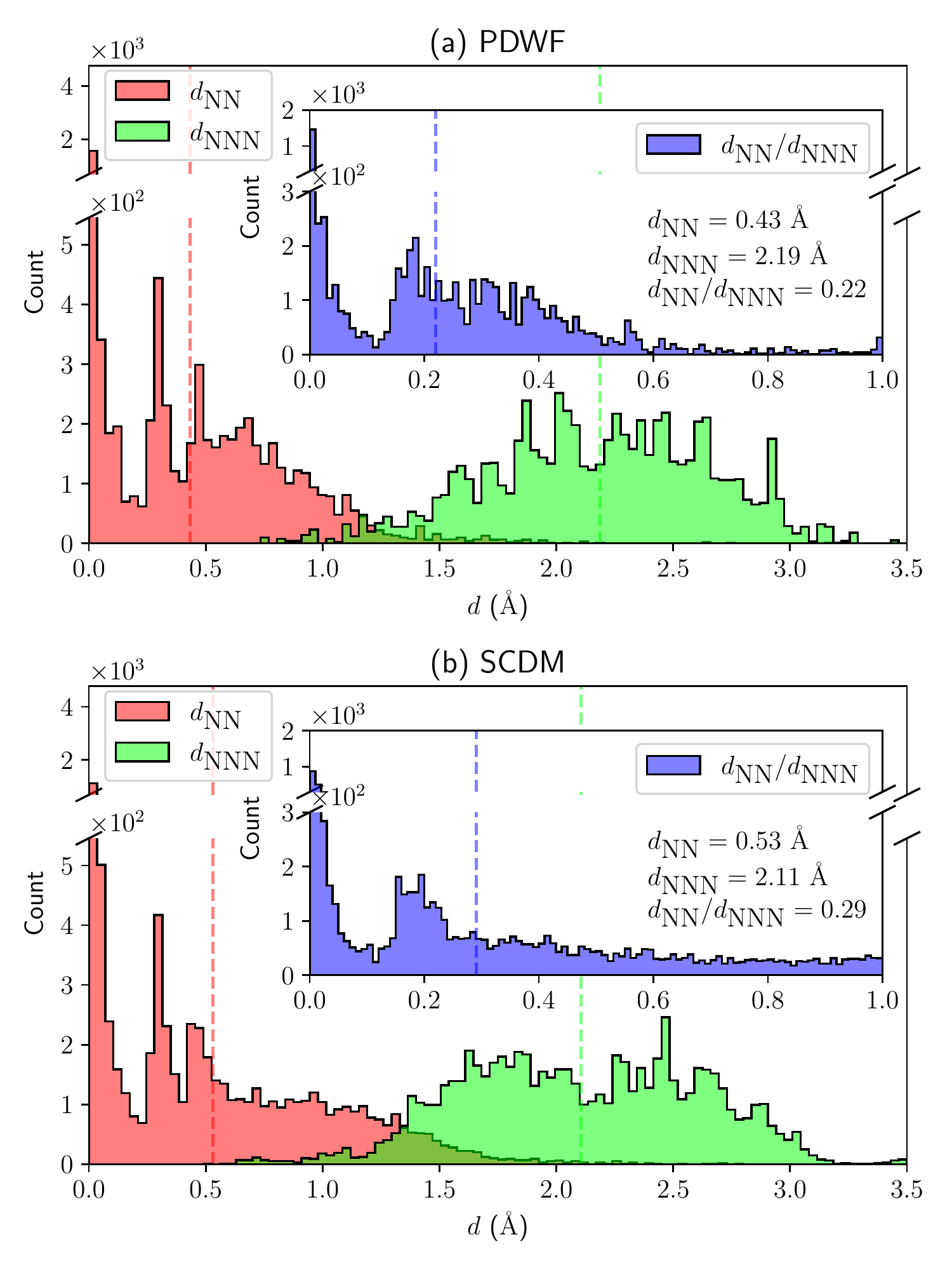}
  \begin{subcaptiongroup}
    \phantomsubcaption\label{fig:center_final_scdm200&nao}
    \phantomsubcaption\label{fig:center_final_scdm200&scdm}
  \end{subcaptiongroup}
  \cprotect\caption{\textbf{Histogram of the distances of the WF centers from
      the NN atom (red, $d_{\text{NN}}$) and NNN atom (green, $d_{\text{NNN}}$),
      for 200 reference structures.}
    \subref{fig:center_final_scdm200&nao} PDWF and
      \subref{fig:center_final_scdm200&scdm} SCDM.
    The inset of each panel shows the histogram of the ratio of $d_{\text{NN}} /
        d_{\text{NNN}}$.
    The numbers in the lower right of each inset are the averages over all the
      \num{6818} MLWFs; PDWF provides MLWFs that are both closer to the NN atom and
      further away from the NNN atom.
  } \label{fig:center_final_scdm200}
\end{figure}

Since we are aiming at restoring a tight-binding atomic-orbital picture with
  PDWF, we compare the distance of the WF centers from the nearest-neighboring
  (NN) and next-nearest-neighboring (NNN) atoms, again both for SCDM and PDWF.
For each method, we compute $d_{\text{NN}}$ and $d_{\text{NNN}}$, \ie, the
  average distance of all the \num{6818} MLWFs from the respective NN and NNN
  atoms.
If $d_{\text{NN}}$ is 0, then the atomic-orbital picture is strictly preserved.
However, this is unlikely to happen since there is no constraint on the WF
  centers during both the disentanglement and the localization, and the final
  PDWFs, resembling atomic orbitals, are optimized according to the chemical
  environment.
Still, if a WF center is much closer to the NN atom than to the NNN atom, then
  one can still assign it to the NN atom, preserving the atomic-orbital picture.
\Cref{fig:center_final_scdm200} shows the histograms for $d_{\text{NN}}$ and
  $d_{\text{NNN}}$ for the two methods.
The PDWF average $d_{\text{NN}}$ = \SI{0.43}{\angstrom} is smaller than the
  SCDM $d_{\text{NN}}$ = \SI{0.53}{\angstrom}, and correspondingly the PDWF
  $d_{\text{NNN}}$ = \SI{2.19}{\angstrom} is instead larger than the SCDM
  $d_{\text{NNN}}$ = \SI{2.11}{\angstrom}.
This can also be observed in \cref{fig:center_final_scdm200}: The overlap of
  the $d_{\text{NN}}$ and $d_{\text{NNN}}$ histograms is smaller for PDWF than
  for SCDM.
To further understand the overlaps, we plot the histogram of the ratio
  $d_{\text{NN}} / d_{\text{NNN}}$ of each MLWF in the insets of
  \cref{fig:center_final_scdm200}.
For a MLWF, if $d_{\text{NN}} / d_{\text{NNN}} = 1$, then the MLWF is a bonding
  orbital centered between two atoms; while if $d_{\text{NN}} / d_{\text{NNN}}
    \ll 1$, then it can be regarded as an (almost) atomic orbital.
The histogram of the ratio of SCDM has a long tail extending towards 1.0, \ie,
  there are a large number of SCDM MLWFs sitting close to bond centers; on the
  contrary, the vast majority of the PDWF MLWFs are closer to the NN atom.

We can further compare the effect of maximal localization on the WF centers.
The WFs from the projection matrices $A_{mn\mathbf{k}}$ are strictly
  atom-centered, \ie $d_{\text{NN}} = 0$.
The inset of \cref{supp-fig:center_scdm200&nn_nao} shows the histogram of the
  initial WFs, \ie, after disentanglement and before maximal localization, and
  the final MLWFs, \ie, after maximal localization, for PDWF.
If one chooses $d_{\text{NN}} \le$ \SI{0.1}{\angstrom} as the criterion for
  atom-centered MLWFs, then $5594/6818$ = 82.0\% of the initial WFs and
  $2045/6818$ = 30.0\% of the final MLWFs are atom-centered.
The disentanglement and maximal localization improve the band interpolation,
  but since there is no constraint on the WF center in the spread functional
  \cref{eq:mv_spreads}, many of the final MLWF centers are not atom-centered.
As a comparison, for SCDM, $955/6818$ = 14.0\% of the initial WFs and
  $1823/6818$ = 26.7\% of the final MLWFs are atom-centered.
For completeness, the statistics and histograms of initial and final
  $d_{\text{NN}}$, $d_{\text{NNN}}$, and $d_{\text{NN}} / d_{\text{NNN}}$ are
  shown in \cref{supp-tab:wfc} and \cref{supp-fig:center_scdm200}.

In summary, for PDWF, most of the initial WFs (after disentanglement and before
  maximal localization) are atom-centered; many drift a bit away from atom
  centers during the localization, but the MLWFs are still much closer to the NN
  than to NNN atoms.
For SCDM, most of the initial WFs are away from atom centers, and maximal
  localization pushes some of the WFs back to atoms, but there is still a large
  number of MLWFs for which an atom representing the WF center cannot be clearly
  identified.
To exactly fix the MLWFs to atomic positions, one needs to add constraints to
  the spread functional \cite{Wang2014}, at the cost of potentially having worse
  interpolators.
However, this is beyond the scope of the current work, and here we rely on the
  atom-centered PAO projectors to guide the MLWFs towards the atomic positions,
  so that the final MLWFs are optimally localized and atom-centered.

\subsubsection{MLWF spreads}
\begin{figure}[tb]
  \includegraphics[width=\linewidth,max height=0.6\textheight]{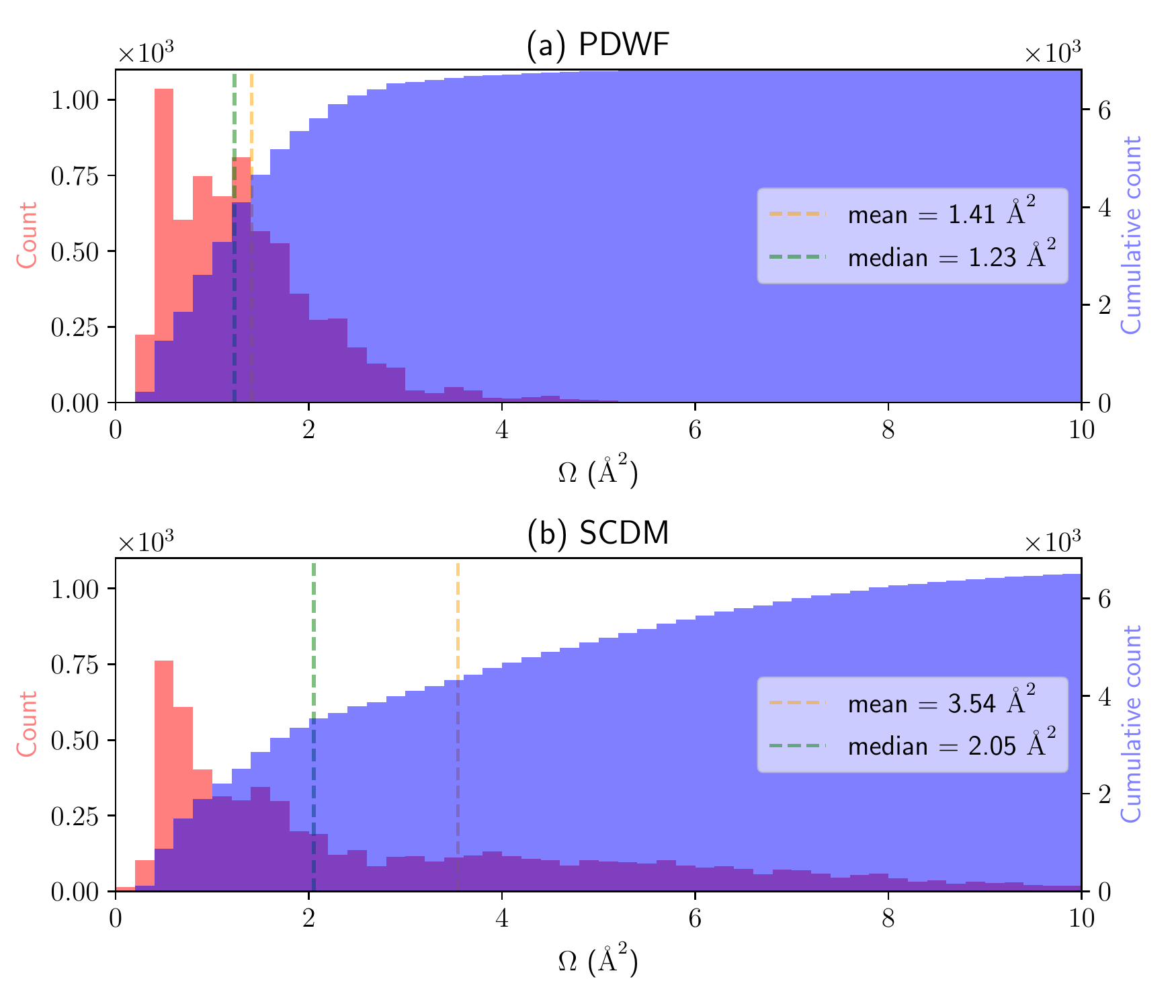}
  \begin{subcaptiongroup}
    \phantomsubcaption\label{fig:sprd_scdm200&nao}
    \phantomsubcaption\label{fig:sprd_scdm200&scdm}
  \end{subcaptiongroup}
  \cprotect\caption{ \textbf{Histogram (red) and
      cumulative histogram (blue) of WF spreads for 200 reference structures.}
    \subref{fig:sprd_scdm200&nao} PDWF and \subref{fig:sprd_scdm200&scdm} SCDM.
    The orange (green) vertical line is the mean (median) spread of the \num{6818}
      MLWFs, their values are shown in the right of each panel.
    The long tail of MLWF spreads obtained with SCDM is absent in PDWF.
  } \label{fig:sprd_scdm200}
\end{figure}

Next, we investigate the spread distributions of SCDM and PDWF.
Usually, we want localized MLWFs to restore the TB atomic orbitals.
\Cref{fig:sprd_scdm200} shows the histograms of the spread distributions for
  the two methods.
The SCDM spreads have a long tail extending over 10 \si{\angstrom^2} in
  \cref{fig:sprd_scdm200&scdm}, due to its inclusion of free-electron states in
  the density matrix, thus resulting in more delocalized MLWFs as discussed
  earlier (see \eg \cref{fig:wf_shape_graphene}).
On the contrary, the PDWF selects and freezes atomic-orbital states from the
  remaining bands, leading to much more localized MLWFs, thus much more clustered
  in a narrow range of \SIrange{0}{4}{\angstrom^2}, and already at
  \SI{5}{\angstrom^2} the cumulative histogram almost reaches the total number of
  MLWFs (see \cref{fig:sprd_scdm200&nao}).
This can be interpreted as follows: The PAO initial projections guide the
  spread minimization toward the (local) minimum resembling spherical harmonics,
  whereas the SCDM-decomposed basis vectors are designed to be mathematical
  objects spanning as much as possible the density matrix, but result in WFs for
  which it is harder to assign definite orbital characters.

We can further compare the average initial (after disentanglement but before
  maximal localization) and final (after disentanglement and maximal
  localization) spreads between the two methods, as shown in
  \cref{supp-tab:spreads} and corresponding histograms in
  \cref{supp-fig:sprd_scdm200_range10_20}.
Maximal localization is needed to bring SCDM spreads, from the initial
  $\Omega^i$ = \SI{30.82}{\angstrom^2} to the final $\Omega^f$ =
  \SI{3.54}{\angstrom^2}; For PDWF, the initial $\Omega^i$ =
  \SI{2.72}{\angstrom^2} is already excellent, and much better than the final
  $\Omega^f$ for SCDM; localization then brings it to an optimal $\Omega^f$ =
  \SI{1.41}{\angstrom^2}.

\subsubsection{Hamiltonian decay}\label{sec:hamiltonian_decay}

\begin{figure}[tb]
  \includegraphics[width=\linewidth,max height=0.6\textheight]{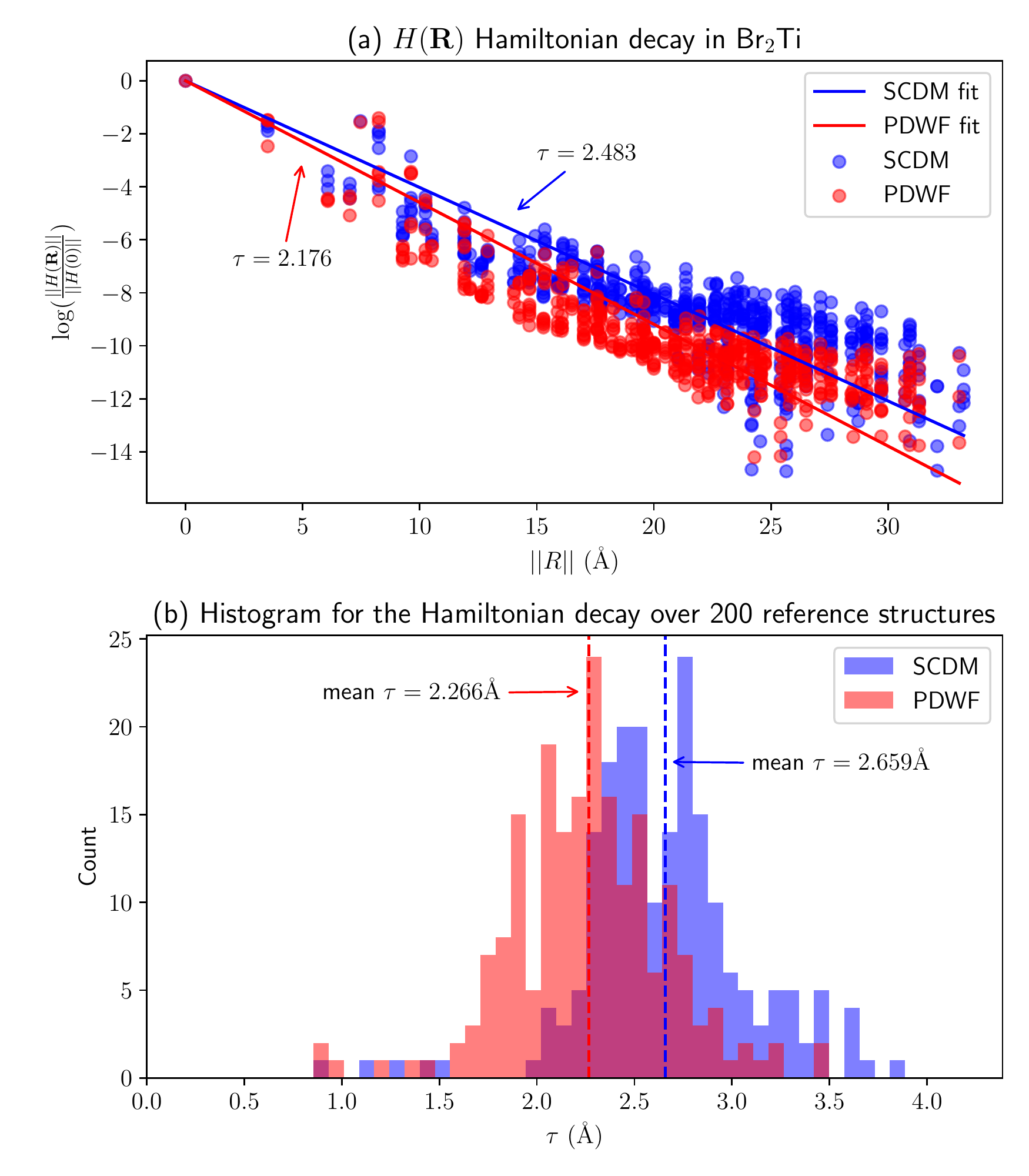}
  \begin{subcaptiongroup}
    \phantomsubcaption\label{fig:hamiltonian&br2ti}
    \phantomsubcaption\label{fig:hamiltonian&hist}
  \end{subcaptiongroup}
  \cprotect\caption{ \textbf{Exponential decay of the Hamiltonian $H(\mathbf{R})$ in the basis of MLWFs.}
    \subref{fig:hamiltonian&br2ti} Exponential-form fitting of Frobenius norm of
      the Hamiltonian $\left\Vert H(\mathbf{R}) \right\Vert$ \wrt to the 2-norm of
      lattice vector $\left\Vert\mathbf{R}\right\Vert$ for the case of \BrtwoTi, for
      PDWF (red) and SCDM (blue).
    The $\tau$ reported are the fitted decay lengths of the PDWF and SCDM
      Hamiltonians, respectively.
    \subref{fig:hamiltonian&hist} Histogram of decay lengths $\tau$ for the 200
      reference materials, obtained using PDWF (red) and SCDM (blue).
    The vertical lines indicate the mean $\tau$ of PDWF and SCDM, respectively.
  } \label{fig:hamiltonian}
\end{figure}

Finally, we compare the decay length of the Wannier gauge Hamiltonian between
  the two methods in \cref{fig:hamiltonian}.
Thanks to the localization of MLWFs, the expectation values of quantum
  mechanical operators in the MLWF basis, such as the Hamiltonian
  $H(\mathbf{R})$, decay rapidly with respect to the lattice vector $\mathbf{R}$
  (exponentially in insulators\cite{Brouder2007,Panati2013} and properly
  disentangled metals).
To compare this decay for the Hamiltonian matrix elements, we approximate the
  Frobenius norm of the Hamiltonian as
  \begin{equation}
    \label{eq:ham_norm}
    \left\Vert H(\mathbf{R}) \right\Vert = \left\Vert H(\mathbf{0}) \right\Vert
    \exp \left(- \frac{ \left\Vert \mathbf{R} \right\Vert }{ \tau }\right),
  \end{equation}
  where $\tau$ measures the decay length.
Then $\tau$ is fitted by least squares to the calculated $\left\Vert
    H(\mathbf{R}) \right\Vert$; as shown in \cref{fig:hamiltonian&br2ti}, the
  Hamiltonian of PDWF decays faster than SCDM for \BrtwoTi, which is selected
  here to represent the general trend between PDWF and SCDM Hamiltonians.
\cref{fig:hamiltonian&hist} shows the histogram of $\tau$ for the 200
  materials; the mean $\tau$ are \SI{2.266}{\angstrom} for PDWF and
  \SI{2.659}{\angstrom} for SCDM, respectively, indicating that the PDWF
  Hamiltonian decays faster than SCDM, consistent with the better band
  interpolation of PDWF discussed in \cref{fig:bandsdist_scdm200}.

\subsection{High-throughput Wannierization}
\begin{figure}[!htb]
  \includegraphics[width=\linewidth,max height=0.6\textheight]{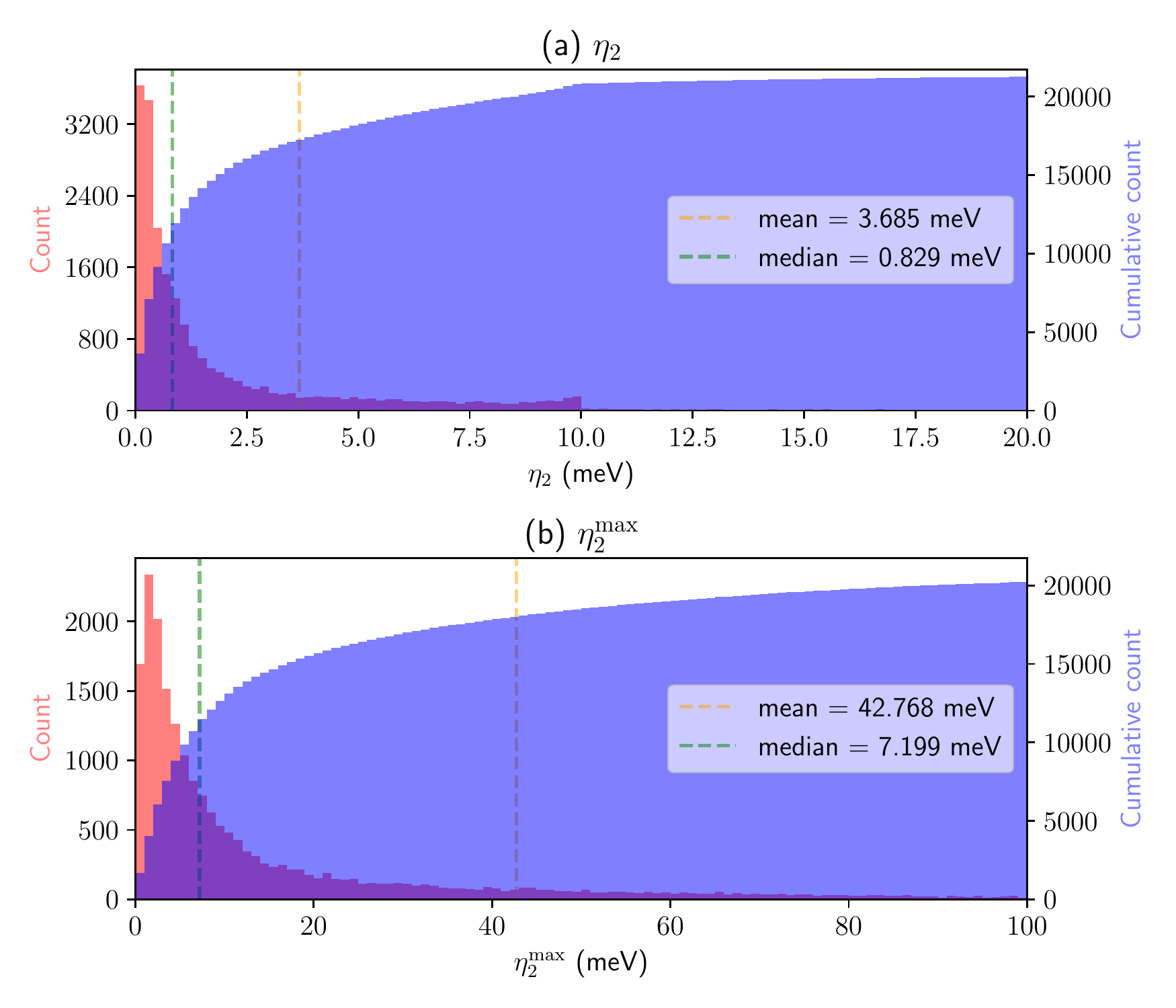}
  \begin{subcaptiongroup}
    \phantomsubcaption\label{fig:bandsdist_lumi&eta}
    \phantomsubcaption\label{fig:bandsdist_lumi&etamax}
  \end{subcaptiongroup}
  \cprotect\caption{\textbf{Histogram (red) and cumulative histogram (blue) of
      the PDWF band distances for \num{21737} non-magnetic structures obtained
      from the materials cloud MC3D database \cite{MC3D}.}
    \subref{fig:bandsdist_lumi&eta} Average band distance $\eta_2$ and
      \subref{fig:bandsdist_lumi&etamax} max band distance $\eta_2^{\max}$.
    The orange (green) vertical line is the mean (median) of the band distance for
      the \num{21737} structures; their values are shown in the right of each panel.
  }
  \label{fig:bandsdist_lumi}
\end{figure}

Based on the above verification, we run a HT Wannierization using PDWF for
  \num{21737} materials, selected from the non-magnetic materials of the MC3D
  database \cite{MC3D}.
\Cref{fig:bandsdist_lumi} shows the band distance histograms for $\eta_2$ and
  $\eta_2^{\max}$.
Overall, the statistics follow the same trend as the 200 materials set in
  \cref{fig:bandsdist_scdm200}: the average $\eta_2$ and average $\eta_2^{\max}$
  are \SI{ 3.685 }{meV} and \SI{ 42.768 }{meV}, respectively.
Note in \cref{fig:bandsdist_lumi&eta} the $\eta_2$ is not truncated at
  \SI{10}{meV}, but rather due to the automated optimization workflow: results
  that have $\eta_2$ larger than a threshold (\SI{10}{meV}) are further optimized
  with respect to \pmin and \pmax, thus improving the average band distance
  $\eta_2$.
In \cref{supp-tab:bandsdist_lumi} we show several other statistics for the band
  distances.
The excellent interpolation quality of PDWF can be assessed, for instance, from
  the number of systems with $\eta_2 \le$ \SI{20}{meV}, that are $\approx 97.8\%$
  of all the calculations ($ 21259 / 21737 $); the corresponding bands distance
  calculated on these {21259} calculations is $\eta_2^{\le 20}$ =
  \SI{2.118}{meV}.
This remarkable result show how automated and reliable Wannierizations can now
  be deployed automatically both for individual calculation and for HT
  application.

\subsection{Additional PAOs for high-energy high-accuracy interpolation}\label{sec:add_paos}
Based on the HT Wannierization results, one can identify cases where the
  interpolation quality can be further improved by increasing the number of PAOs.
Typically, the number of PAOs is determined during pseudopotential generation,
  and they are usually the orbitals describing low-energy valence electrons.
In some cases, the bonding/anti-bonding combinations of these PAOs are not
  sufficient to span the space of target conduction bands, leading to a loss of
  interpolation quality.
We use silicon as an example to illustrate the difficulties of accurately
  describing its CBM \cite{Ponce2021}, which is not located at any high-symmetry
  $k-$point, but along the $\Gamma-$X line.
The common choice of one $s$ and three $p$ hydrogenic or PAOs projectors per
  atom results in oscillations in the Wannier-interpolated bands at the meV
  level.
To remedy this, one can use a larger set of PAOs, \eg, by regenerating a
  silicon pseudopotential including $d$ PAOs as discussed in \cref{sec:silicon}.
However, generating a new pseudopotential requires extensive testing and
  validation, therefore another solution could be using a set of PAOs different
  from the pseudopotential ones.
To compare this second approach, we test here also PAOs obtained from the \OMX
  code \cite{Ozaki2004}, and Wannierize silicon using one $s$, three $p$, and
  five $d$ PAOs per atom using ED.
This provides a much better description of the CBM, as shown in
  \cref{supp-fig:si_bands_cbm} Moreover, the additional $d$ orbitals allow to
  raise the inner energy window and better reproduce a larger number of
  conduction bands, as shown in \cref{supp-fig:si_bands_openmx}, which might be
  beneficial for some applications. %
For completeness, we also show the WF spreads and shapes of $d$ orbitals in
  \cref{supp-fig:wf_shape_silicon_openmx}.
However, there are some caveats to this approach.
When using external PAOs, ideally one should generate them using the same
  pseudization scheme as the pseudopotentials used in the DFT calculations.
The PAOs from \OMX are instead generated using a different scheme, resulting in
  lower projectabilities (smaller than one even for the valence bands, as shown
  in \cref{supp-fig:si_projectability_openmx_custom_pseudo}).
In such case, PD cannot reproduce the original bands (see
  \cref{supp-fig:si_bands_openmx_custom_pseudo&omx_pd}), thus ED (with a higher
  inner energy window) is needed to obtain accurate interpolation (see
  \cref{supp-fig:si_bands_openmx&omx_spd_wd10}).
In comparison, the pseudopotential PAOs which we regenerated with 3$d$ orbitals
  (as discussed in \cref{sec:silicon}) are better projectors for the
  wavefunctions.
Indeed, the first 12 bands have projectabilities almost equal to 1, and as a
  consequence PD itself already provides accurate band interpolation (all the
  low-energy conduction states are frozen since their projectabilities are high,
  see \cref{supp-fig:si_bands_openmx_custom_pseudo&qe_pd}).
Moreover, we mention that when adding additional projectors one needs to make
  sure that they have the correct number of radial nodes: \eg, the gold
  pseudopotential from SSSP \cite{Prandini2018} contains $5s+5p$ semicore states,
  and $6s+5d$ orbitals for valence electrons.
If one wants to add an additional $6p$ orbital, it is important to ensure that
  the $6p$ orbital has one radial node, such that it is orthogonal to the
  nodeless $5p$ semicore state; Otherwise, the Bloch wavefunctions would project
  onto the $5p$ semicore state, and PD would only disentangle the $5p$ semicore
  states instead of the $6p$ orbitals contributing to bands above the Fermi
  energy.
In summary, including more projectors can further improve the interpolation
  quality, but at the expense of increasing the number of orbitals in the model.
The combination of PD and ED enables to improve the interpolation quality of
  low-projectability states while keeping the TB model size small.
Automatic checks could be implemented in the future in the \AiiDA workflows to
  detect whether the projectability drops below a certain threshold, and in that
  case either raise a warning or automatically add more projectors.

\section{Conclusions}
We present an automated method for the automated, robust, and reliable
  construction of tight-binding models based on MLWFs.
The approach applies equally well to metals, insulators and semiconductors,
  providing in all cases atomic-like orbitals that span both the occupied states,
  and the empty ones whose character remains orbital-like and and not
  free-electron-like.
The method is based on the band projectability onto pseudo-atomic orbitals to
  select which states are kept identically, dropped, or passed on to the
  established disentanglement procedure.
We augment such projectability-based selection with an additional energy window
  to guarantee that all states around the Fermi level or the conduction band edge
  are well reproduced, showing that such a combination enables accurate
  interpolation even when minimal sets of initial atomic orbitals are chosen.
This results in compact Wannier tight-binding models that provide accurate band
  interpolations while preserving the picture of atomic orbitals in crystals.
We refer to the method collectively as projectability-disentangled Wannier
  functions (PDWF).

The Wannierization process is implemented as fully automated \AiiDA workflows.
We compare PDWFs with the other method that is also fully automated, namely
  SCDM.
We show with a detailed study of 200 structures that PDWFs lead to more
  accurate band interpolations (with errors with respect to the original bands at
  the \si{meV} scale), and are more atom-centered and more localized than those
  originating from SCDM.
The high accuracy in band interpolations, the target atomic orbitals obtained,
  and the low computational cost make PDWFs an ideal choice for automated or
  high-throughput Wannierization, which we demonstrate by performing the
  Wannierization of \num{21737} non-magnetic structures from the Materials Cloud
  MC3D database.

\section{Methods}

We implement the PAO projection in the \texttt{pw2wannier90.x} executable
  inside \QE (QE) \cite{Giannozzi2009,Giannozzi2020a}; the PD and PD+ED methods
  are implemented on top of the \WAN code \cite{Pizzi2020}.
In terms of the practical implementation, computing PAO projections is more
  efficient in both computational time and memory than the SCDM QR decomposition
  with column pivoting (QRCP) algorithm, since the $A_{mn\mathbf{k}}$ matrices
  (\ie, the inner products of Bloch wavefunctions with PAOs) can be evaluated in
  the plane-wave $G$ vector space, rather than requiring a Fourier transform and
  decomposition of very large real-space wavefunction matrices.
Furthermore, since the HT Wannierization can be computationally intensive, we
  implement a ``$k-$pool parallelization strategy'' inside \PtoW, similarly to
  the main \texttt{pw.x} code of QE, to efficiently utilize many-core
  architectures by parallelizing over ``pools'' of processors for the almost
  trivially-parallel computations at each $k-$point.
Test results show that $k-$pool parallelization significantly improves the
  efficiency of \PtoW (benchmarks are shown in \cref{supp-fig:p2w_kpool}).

The DFT calculations are carried out using QE, with the SSSP efficiency
  (version 1.1, PBE functional) library \cite{Prandini2018} for pseudopotentials
  and its recommended energy cutoffs.
The HT calculations are managed with the \AiiDA infrastructure
  \cite{Pizzi2016,Huber2020,Uhrin2021} which submits QE and \WAN calculations to
  remote clusters, parses, and stores the results into a database, while also
  orchestrating all sequences of simulations and workflows.
The automated \AiiDA workflows are open-source and hosted on \texttt{GitHub}
  \cite{aiidaW90}.
The workflows accept a crystal structure as input and provide the
  Wannier-interpolated band structure, the real-space MLWFs, and a number of
  additional quantities as output.
Semicore states from pseudopotentials are automatically detected and excluded
  from the Wannierizations, except for a few cases where some semicore states
  overlap with valence states; in such cases, all the semicore states are
  Wannierized, otherwise the band interpolation quality would be degraded,
  especially for SCDM.
A regular $k-$point mesh is used for the Wannier calculations, with a $k-$point
  spacing of \SI{0.2}{\per\angstrom}, as selected by the protocol in
  \citet{Vitale2020}.
MLWFs are rendered with VESTA \cite{Momma2011}.
Figures are generated by \texttt{matplotlib} \cite{Hunter2007}.

\section{Data Availability}
All data generated for this work can be obtained from the Materials Cloud
  Archive (\url{https://doi.org/10.24435/materialscloud:v4-e9}).

\section{Code Availability}

All codes used for this work are open-source; the latest stable versions can be
  downloaded at \url{http://www.wannier.org/},
  \url{https://www.quantum-espresso.org/}, \url{https://www.aiida.net/}, and
  \url{https://github.com/aiidateam/aiida-wannier90-workflows}.

The modifications to the codes mentioned above implemented for this work will
  become available in the next releases of \QE (\PtoW) and \WAN.

\section{Acknowledgements}
We acknowledge financial support from the NCCR MARVEL (a National Centre of
  Competence in Research, funded by the Swiss National Science Foundation, grant
  No.
205602), the Swiss National Science Foundation (SNSF) Project Funding
(grant 200021E\smallunderscore{}206190 ``FISH4DIET'').
The work is also supported by a pilot access grant from the Swiss National
  Supercomputing Centre (CSCS) on the Swiss share of the LUMI system under
  project ID ``PILOT MC EPFL-NM 01'', a CHRONOS grant from the CSCS on the Swiss
  share of the LUMI system under project ID ``REGULAR MC EPFL-NM 02'', and a
  grant from the CSCS under project ID s0178.

\section{Author Contributions}
J.~Q.
implemented and tested the PDWF method.
N.~M.
suggested to use projectability thresholds.
G.~P.
and
N.~M.
supervised the project.
All authors analyzed the results and contributed to writing the manuscript.

\section{Competing Interests}
The authors declare that there are no competing interests.

\bibliography{main}

\end{document}